\documentclass[12pt,a4paper]{article}
\usepackage{xcolor}
\usepackage[pdftex,colorlinks=true,hypertexnames=false]{hyperref}
\definecolor{darkblue}{rgb}{0,0,.6}
\hypersetup{citecolor=darkblue,linkcolor=darkblue,urlcolor=darkblue}
\usepackage{amssymb,enumitem,booktabs,subfig,microtype,setspace,bm,longtable,xr,dsfont,orcidlink}
\usepackage[top=1.1in, bottom=1.1in, left=1.1in, right=1.1in]{geometry}
\usepackage{natbib}
\usepackage{url}
\usepackage{amsmath}
\usepackage{amsfonts}
\usepackage{epsfig}
\usepackage{graphics}
\usepackage[normalem]{ulem}
\usepackage{palatino,eulervm}
\usepackage{graphicx}
\usepackage{lscape}
\usepackage{rotating}
\usepackage[linewidth=1pt]{mdframed}
\allowdisplaybreaks[4]

\providecommand{\U}[1]{\protect\rule{.1in}{.1in}}

\graphicspath{{plots/}}
\newsavebox\CBox
\def\textBF#1{\sbox\CBox{#1}\resizebox{\wd\CBox}{\ht\CBox}{\textbf{#1}}}

\usepackage{amsthm,thmtools}
\usepackage{mathrsfs}

\makeatletter
\def\th@newremark{\th@remark\thm@headfont{\bfseries}}
\makeatletter
\theoremstyle{newremark}

\declaretheoremstyle[
  spaceabove=6pt, spacebelow=6pt,
  headfont=\bfseries,
  notefont=\mdseries, notebraces={(}{)},
bodyfont=\normalfont,
  postheadspace=0.5em,
]{mystyle}

\makeatletter
\newcommand*{\addFileDependency}[1]{
\typeout{(#1)}
\@addtofilelist{#1}
\IfFileExists{#1}{}{\typeout{No file #1.}}
}\makeatother

\newcommand*{\myexternaldocument}[1]{%
\externaldocument{#1}%
\addFileDependency{#1.tex}%
\addFileDependency{#1.aux}%
}

\myexternaldocument{supplement}
\newcommand{\Rlogo}{\protect\includegraphics[height=1.8ex,keepaspectratio]{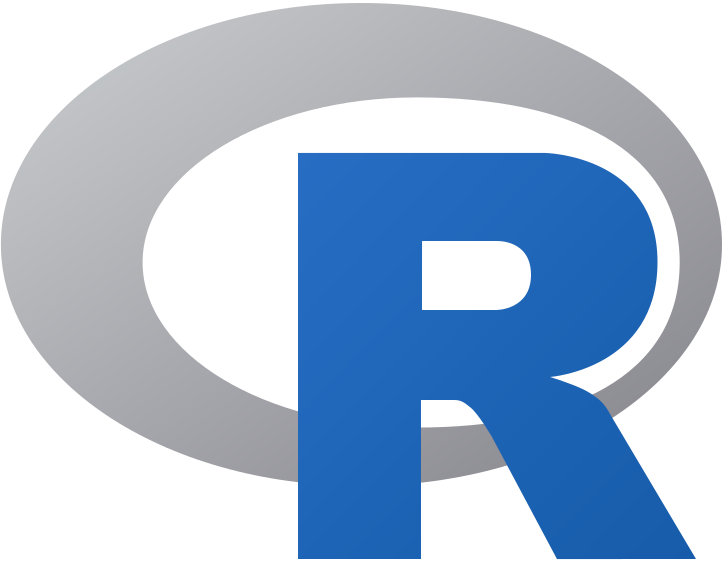}}

\begin{document}

\title{Forecasting Age Distribution of Deaths: \mbox{Cumulative Distribution Function Transformation}}
\author{Han Lin Shang \orcidlink{0000-0003-1769-6430} \footnote{Postal address: Department of Actuarial Studies and Business Analytics, 4 Eastern Road, Macquarie University, NSW 2109, Australia; Email: hanlin.shang@mq.edu.au; Telephone: +61(2) 9850 4689.} \\
Department of Actuarial Studies and Business Analytics \\
Macquarie University \\
\\
Steven Haberman \orcidlink{0000-0003-2269-9759}\\
Bayes Business School \\
City St George's, University of London}

\date{}

\maketitle

\centerline{\bf Abstract}

\medskip

Like density functions, period life-table death counts are nonnegative and have a constrained integral, and thus live in a constrained nonlinear space. Implementing established modelling and forecasting methods without obeying these constraints can be problematic for such nonlinear data. We introduce cumulative distribution function transformation to forecast the life-table death counts. Using the Japanese life-table death counts obtained from the \cite{JMD24}, we evaluate the point and interval forecast accuracies of the proposed approach, which compares favourably to an existing compositional data analytic approach. The improved forecast accuracy of life-table death counts is of great interest to demographers for estimating age-specific survival probabilities and life expectancy and actuaries for determining temporary annuity prices for different ages and maturities.

\medskip


\newpage



\section{Introduction}\label{sec1}
\renewcommand{\theequation}{1.\arabic{equation}}
\setcounter{equation}{0}

Density-valued objects are common, examples of which include income distribution \citep{KU01}, financial return distribution \citep{KMP+19},  distributions of the times when bids are submitted in an online auction \citep{JSP+08}, bike sharing network in Paris \citep{Jiang23}, and age distribution of deaths in demography \citep{SH20}, to name only a few. 


In demography, the age distribution of deaths provides important insights into longevity and lifespan variability that cannot be grasped directly from either the central mortality rate or the survival function. For modelling the age distribution of deaths, \cite{Oeppen08} demonstrated that using compositional data analysis (CoDa) to forecast the age distribution of death counts does not lead to more pessimistic results than forecasting age-specific mortality. The pessimistic mortality rate forecasts may also be due to the linear extrapolation (constant trend) in the well-known Lee-Carter model; in that case, the mortality rate forecasts are higher than the actual ones. Thus, the age distribution of death counts is more suitable than the central mortality rates for computing period life expectancy and annuity values. In addition to providing an informative description of the mortality experience of a population, the age-at-death distribution yields readily available information on the ``central longevity indicators" (e.g., mean, median and modal age at death, see \cite{CRT+05} and \cite{CR10}) and lifespan variability \citep[e.g.,][]{Robine01, VZV11, VC13, VMM14, AV08}.

To model and forecast a time series of density-valued curves, \cite{BCO+07} have put forward a principal component approach to forecast life-table death counts within a CoDa framework by treating life-table death count ($d_x$) for each age $x$ as compositional data. The data are constrained to vary between two limits (e.g., 0 and a constant upper bound), which in turn imposes constraints upon the variance-covariance structure of the raw data. To remove these constraints, a standard approach involves breaking the sum constraint using a transformation of the raw data before applying conventional statistical techniques to the transformed data. The most popular transformation is the centred log-ratio (clr) transformation dated back to pioneering work by \cite{AS80} and \cite{Aitchison82, Aitchison86}. 

An issue with the clr transformation in the CoDa framework is the possible presence of zero counts. Although some ad-hoc ways exist of handling the zero counts \citep[see, e.g.,][]{MBP13, FFM00}, we introduce a novel transformation that naturally handles zero counts. In our method, we first convert an age distribution of death for each year to a probability by dividing its radix $10^5$. So, the life-table death count at each age can be seen as a composition that sums to one. Via the cumulative sum, we transform a probability distribution function into a cumulative distribution function (CDF). In doing so, it requires a fewer number of constraints than the life-table death counts, with an additional benefit of monotonicity \citep[see also][]{MS13}. The inverse of CDF is quantile, which is a key quantity in Wasserstein distance for measuring discrepancy between two distributions \citep[see, e.g.,][]{DM22}. With a time series of CDFs, we model its pattern via a logistic transformation. Within this unconstrained space, we apply a functional time series forecasting method to obtain the $h$-step-ahead curve prediction for a chosen horizon $h$. Taking the inverse logistic transformation, we obtain the $h$-step-ahead life-table death count forecasts after first-order differencing.


Using the Japanese age- and sex-specific life-table death counts from 1975 to 2022, we evaluate and compare one- to 16-step-ahead point and interval forecast accuracy between the CoDa and CDF transformations. To evaluate point forecast accuracy, we use the Kullback-Leibler divergence and the Jensen-Shannon divergence. To assess the interval forecast accuracy, we utilise the interval score of \cite{GR07} and \cite{GK14}; see Section~\ref{sec:4} for details. The improved forecast accuracy of life-table death counts is of great importance to demographers and actuaries for determining remaining age-specific life expectancies and to actuaries for pricing temporary annuities for various ages and maturities.

The remainder of this paper is organised as follows: Section~\ref{sec:2} describes the Japanese data set obtained from the \cite{JMD24}, which is the age- and sex-specific life-table death counts in Japan from 1975 to 2022. Section~\ref{sec:3} introduces the CDF transformation for producing the point and interval forecasts of the age distribution of life-table death counts. Using the point and interval forecast errors in Section~\ref{sec:4}, we evaluate and compare the point and interval forecast accuracies among the methods considered. Section~\ref{sec:5} applies the CDF transformation to estimate the single-premium temporary immediate annuity prices for various ages and maturities for female and male policyholders residing in Japan. Conclusions are presented in Section~\ref{sec:6}, along with some ideas on how the method presented here can be further extended.

\section{Age- and sex-specific life-table death counts in Japan}\label{sec:2}

In many developed nations, such as Japan, increases in longevity risk and an aging population has led to concerns regarding the sustainability of the pension, health and aged care systems \citep[see, e.g.,][]{Coulmas07, OECD20}. Japan has one of the highest average life expectancies in the world, with exceptional longevity in Okinawa \citep{Poulain11}. Japan is also one of the countries that have freely available subnational mortality data, alongside Australia, Canada, France and the US.

\begin{figure}[!htb]
\centering
\includegraphics[width=8.5cm]{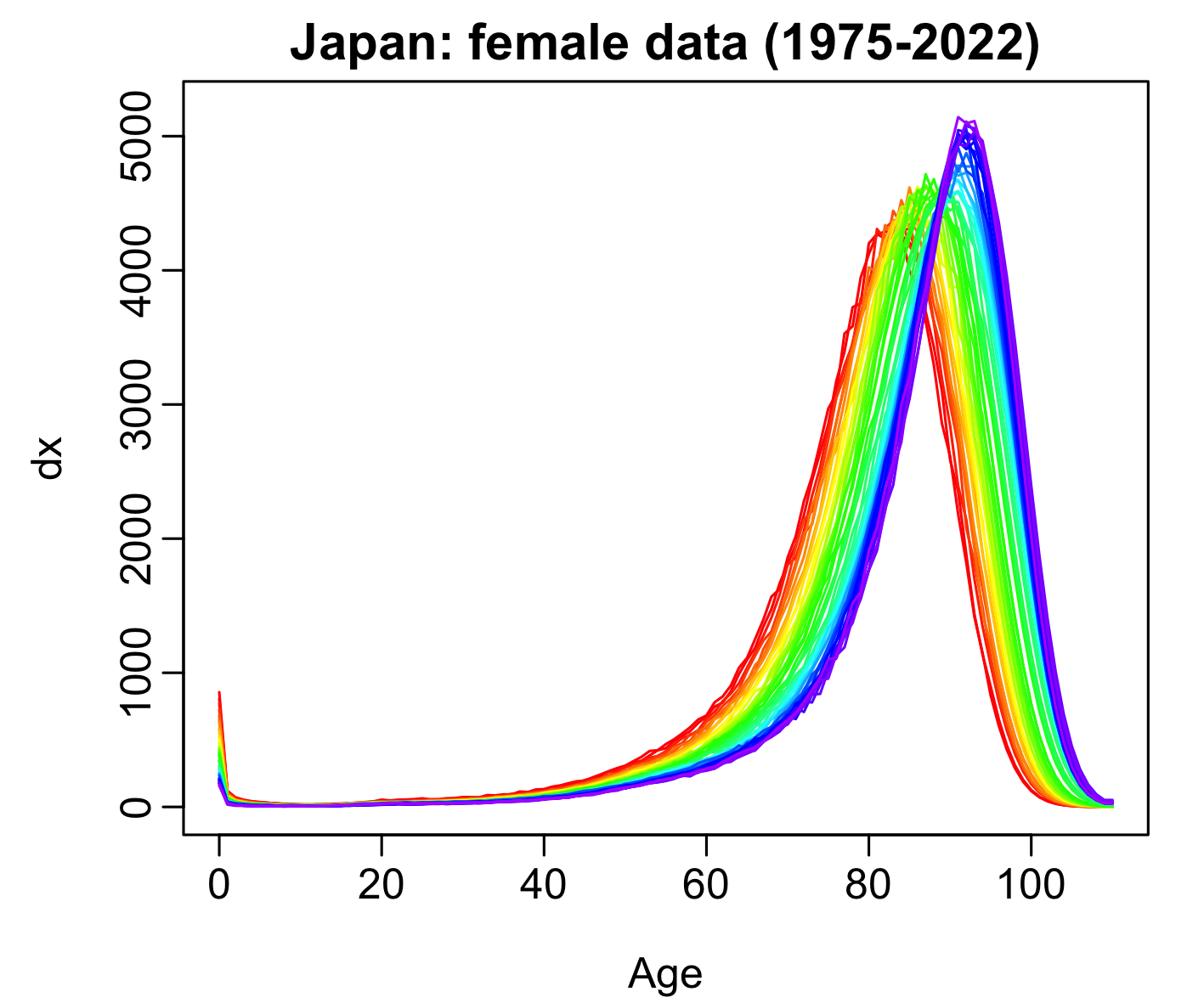}
\quad
\includegraphics[width=8.5cm]{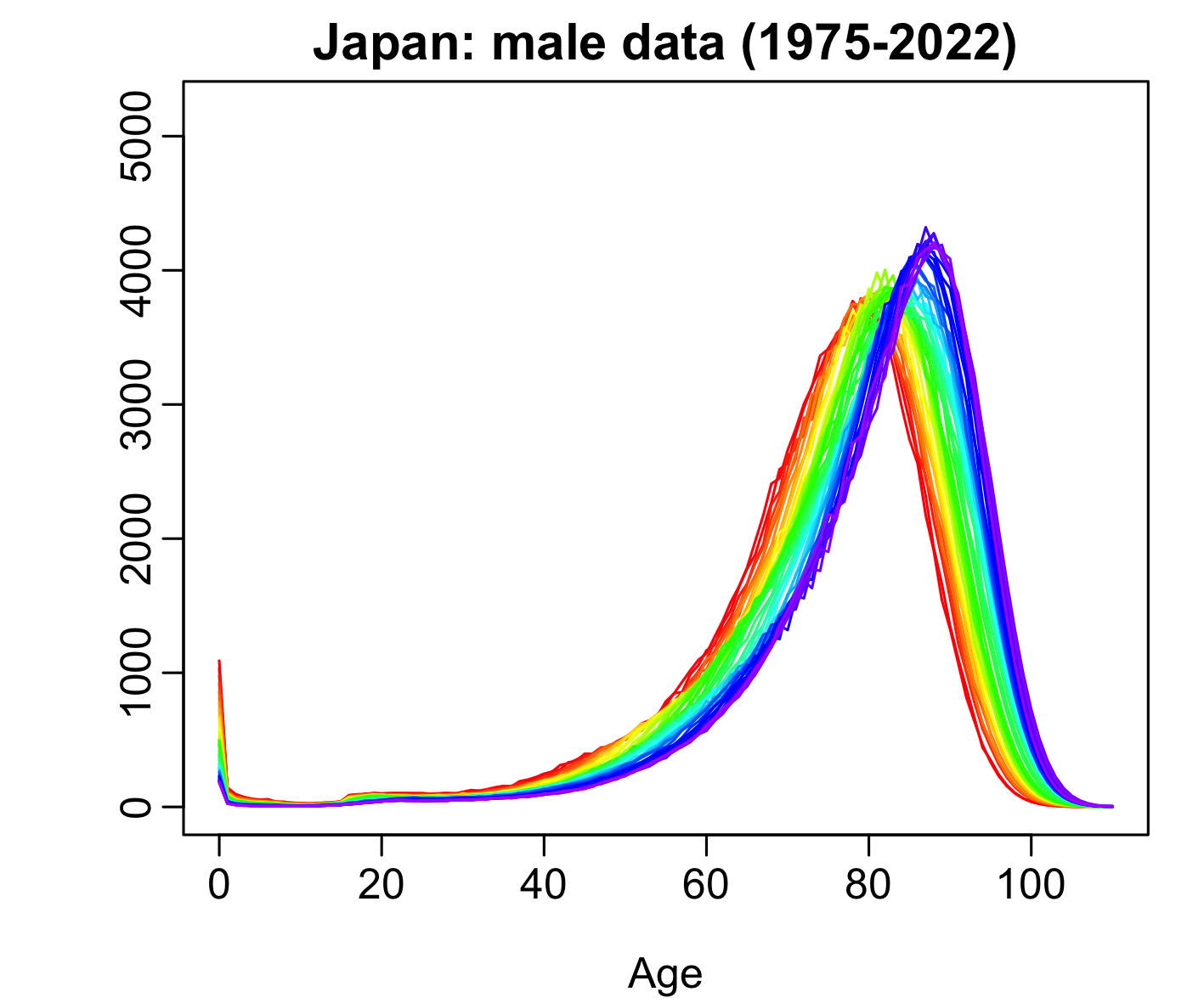}
\caption{\small{Rainbow plots of the age distribution of deaths from 1975 to 2022 in a single-year group. The life-table radix is 100,000 for each year. The life-table death counts in the oldest years are shown in red, while the most recent years are in violet. Curves are ordered chronologically according to the rainbow colours.}}\label{fig:1}
\end{figure}

In Figure~\ref{fig:1}, we display Japanese age- and sex-specific life-table death counts from 1975 to 2022, obtained from the \cite{JMD24}. We use data from the period after the First and Second World Wars to obtain a more stable parameter estimate from the historical data. We study life-table death counts, where the life-table radix (i.e., a population experiencing 100,000 births annually) is fixed at 100,000 at age 0 while the remaining number of people alive is 0 in the last age group for each year. For the life-table death counts, there are 111 ages, and these are ages 0, 1, \dots, 109, 110+. Due to rounding, there are potentially zero counts for age 110+ at some years. To overcome this problem, we work with the probability of dying (i.e., $q_x$) and the life-table radix to recalculate our estimated death counts (up to six decimal places). In doing so, we obtain more precise death counts than the ones reported in the \cite{JMD24}.

Figure~\ref{fig:1} demonstrates a decreasing trend in infant death counts and a typical negatively skewed distribution for the life-table death counts, where the peaks shift to higher ages for both sexes. This shift is a main driver of longevity risk, which is a major issue for insurers and pension funds, especially in the selling and risk management of annuity products \citep[see][for a discussion]{DDG07}. In demography, it becomes increasingly popular to model and forecast \textit{period} life-table death counts. By modelling the life-table death counts, we can understand a redistribution of life-table death counts, where deaths at younger ages gradually shift towards older ages.

\begin{figure}[!htb]
\centering
\includegraphics[width=8.4cm]{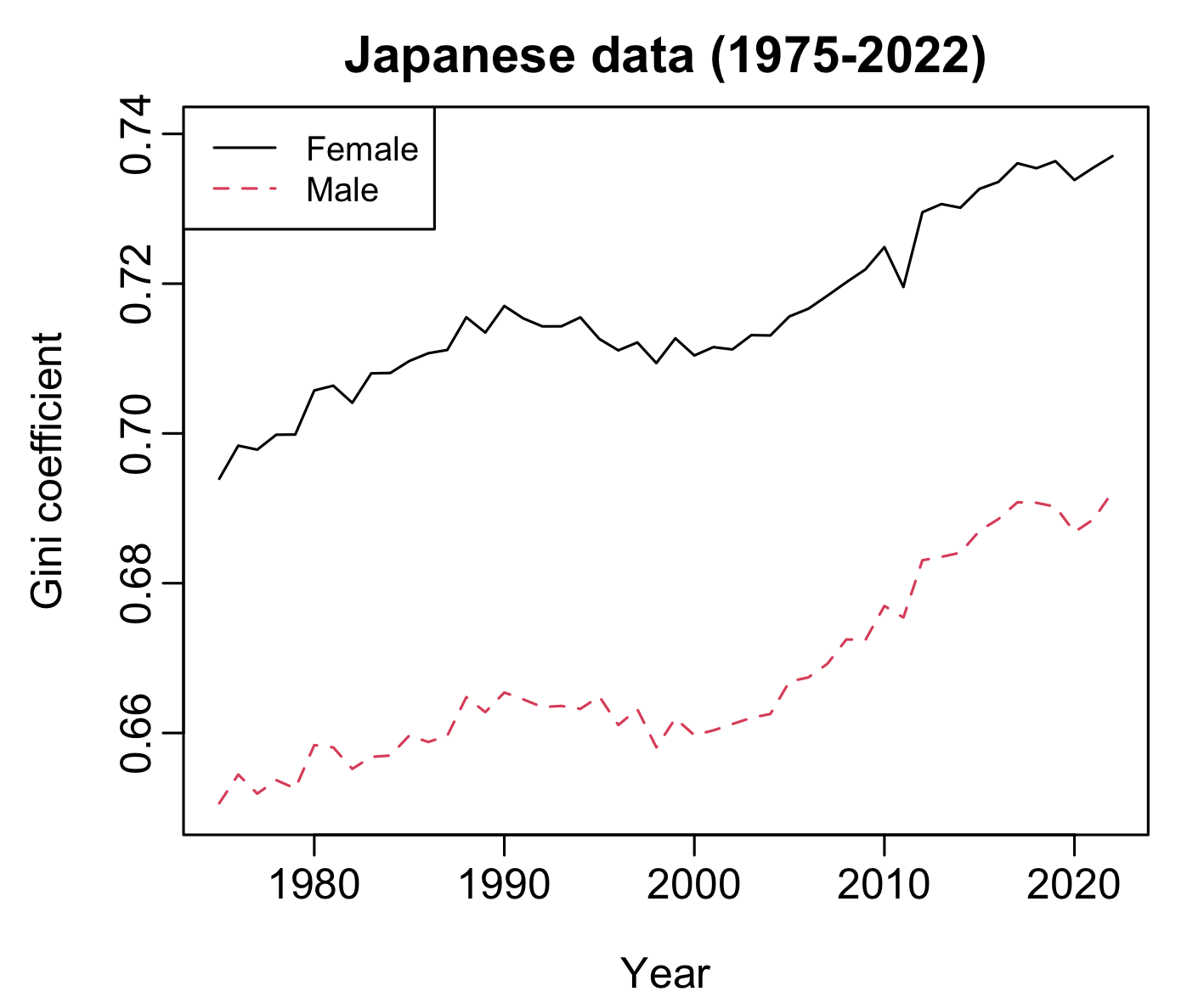}
\caption{\small{Gini coefficients for Japanese age- and sex-specific life-table death counts from 1975 to 2022. When the Gini coefficient approaches 0, it indicates perfect inequality across ages in the life-table death counts. When the Gini coefficient approaches 1, it indicates perfect equality.}\label{fig:2}}
\end{figure}

Lifespan variability can be assessed using measures like the interquartile range or the Gini coefficient \citep[see][]{WH99, VC13, DCH+17}. In life-table death counts, a Gini coefficient of 0 represents perfect inequality among ages, while 1 indicates perfect equality, with deaths occurring at the same age. Figure~\ref{fig:2} shows higher Gini coefficients for females, reflecting greater inequality among males. While not our focus, quantifying and examining temporal changes in distribution spread is a key advantage of modeling age distributions of deaths.

\section{Forecasting methods for density-valued objects}\label{sec:3}

Let us denote age-specific life-table death counts as $d_{t,i}^{s}$, where $t$ denotes a year, $s$ denotes either the female or male series, and $i$ denotes an age. For each year $t$, the life-table death counts sum to a radix~$10^5$. From the viewpoints of proportions and CDFs, it is convenient to scale the life-table death counts to sum to one.

\subsection{Cumulative distribution function transformation}\label{sec:3.1}

We describe the CDF transformation in the following steps:
\begin{enumerate}
\item[1)] Compute the empirical cumulative distribution function via cumulative sum,
\begin{equation*}
D^{s}_{t,x} = \sum_{i=1}^{x}d^{s}_{t,i},  \qquad x=1, \dots, 111, \quad t=1, \dots, n,
\end{equation*}
where $D^{s}_{t,111}=1$ and $s$ represents female or male. In Figure~\ref{fig:3}, we display a time series of the empirical CDFs of the Japanese data from 1975 to 2022.
\begin{figure}[!htb]
\centering
\includegraphics[width=8.5cm]{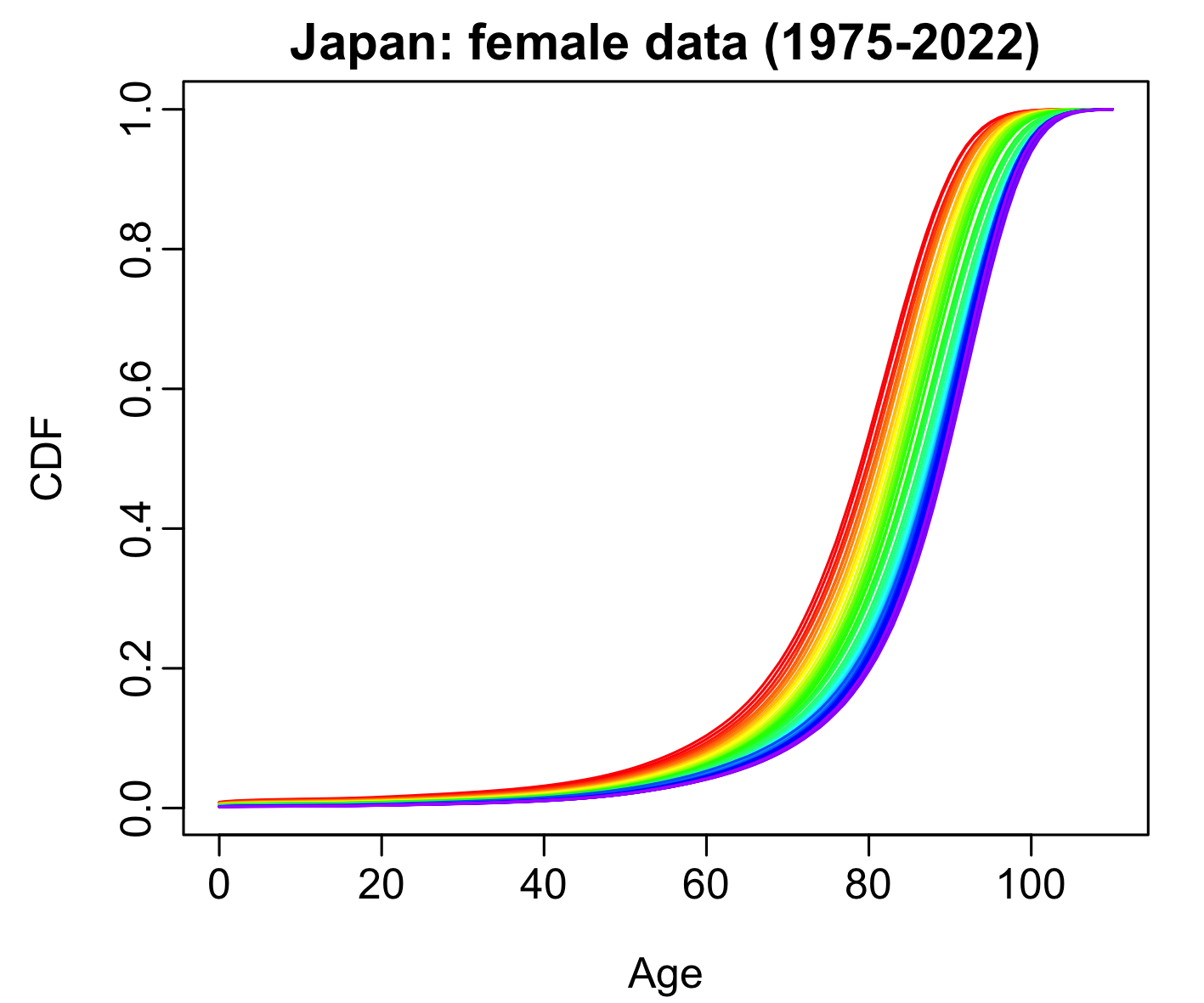}
\quad
\includegraphics[width=8.5cm]{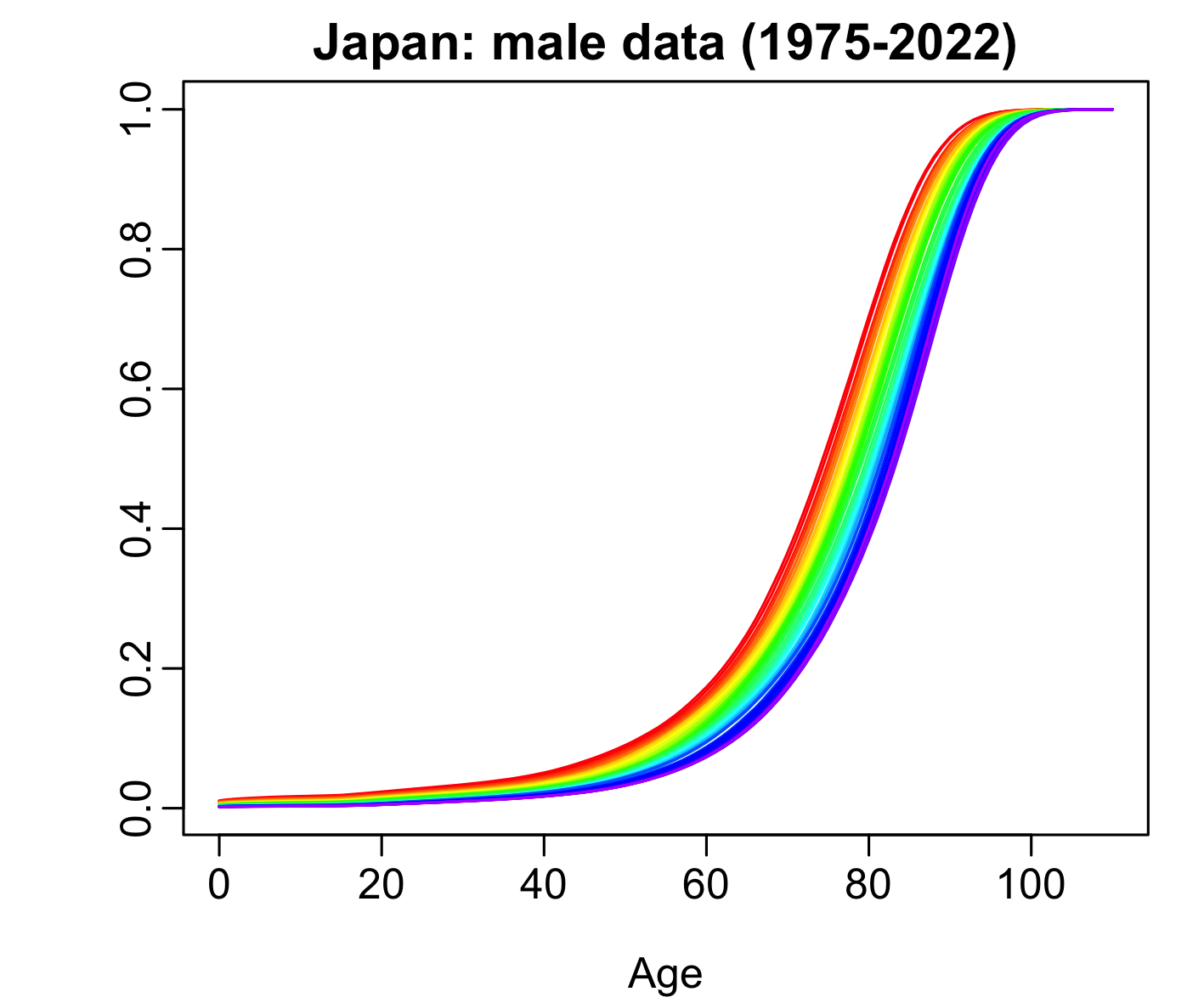}
\caption{\small{Empirical CDFs of the Japanese age- and sex-specific life-table death counts from 1975 to 2022.\label{fig:3}}}
\end{figure}
\item[2)] Perform the logistic transformation,
\begin{equation*}
Z^{s}_{t,x} = \text{logit}(D^{s}_{t,x}) = \ln\Big(\frac{D^{s}_{t,x}}{1-D^{s}_{t,x}}\Big),
\end{equation*}
where $\ln(\cdot)$ denotes the natural logarithm. In Figure~\ref{fig:4}, we present the logit transformation of the empirical CDFs.
\begin{figure}[!htb]
\centering
\includegraphics[width=8cm]{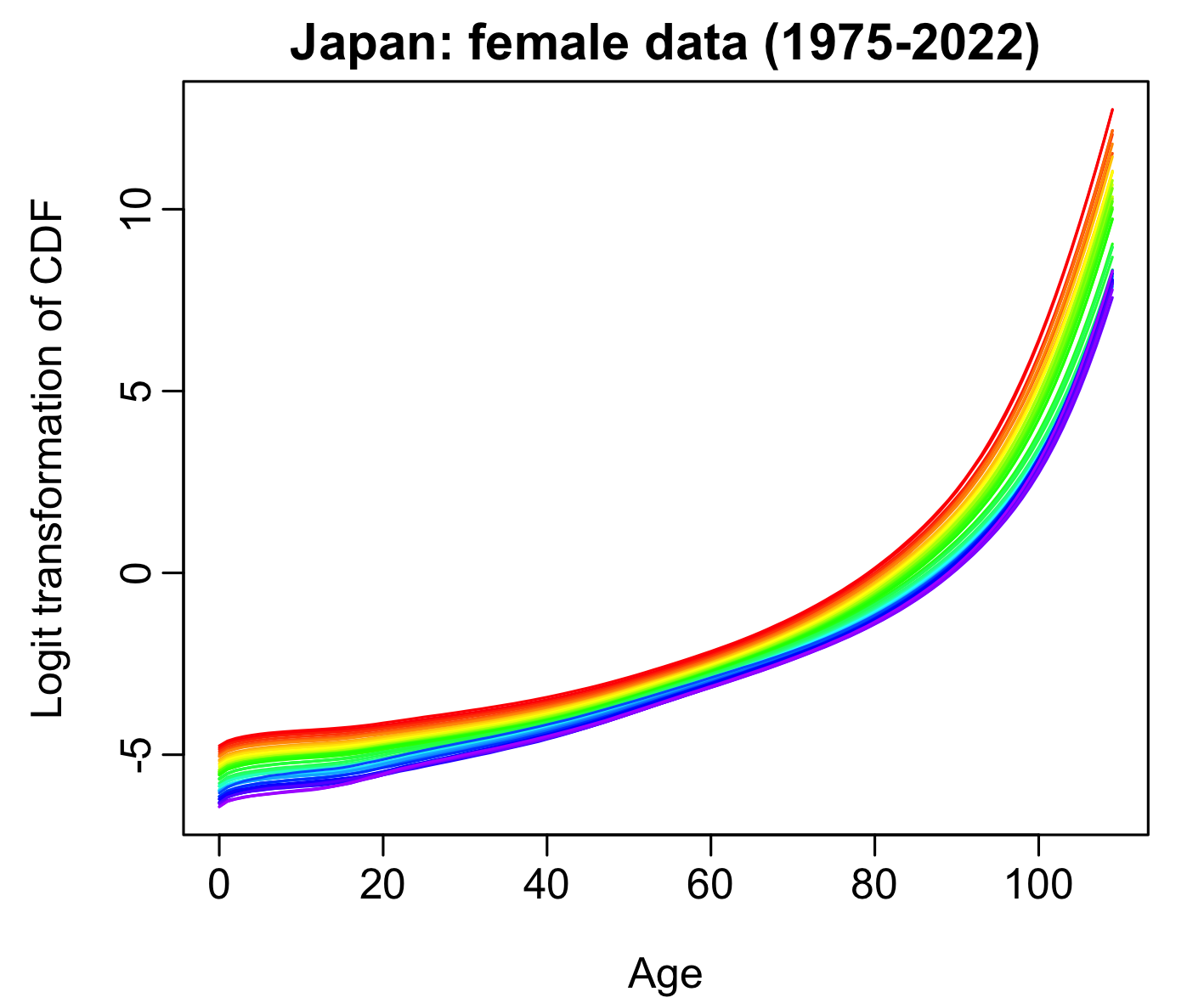}
\quad
\includegraphics[width=8cm]{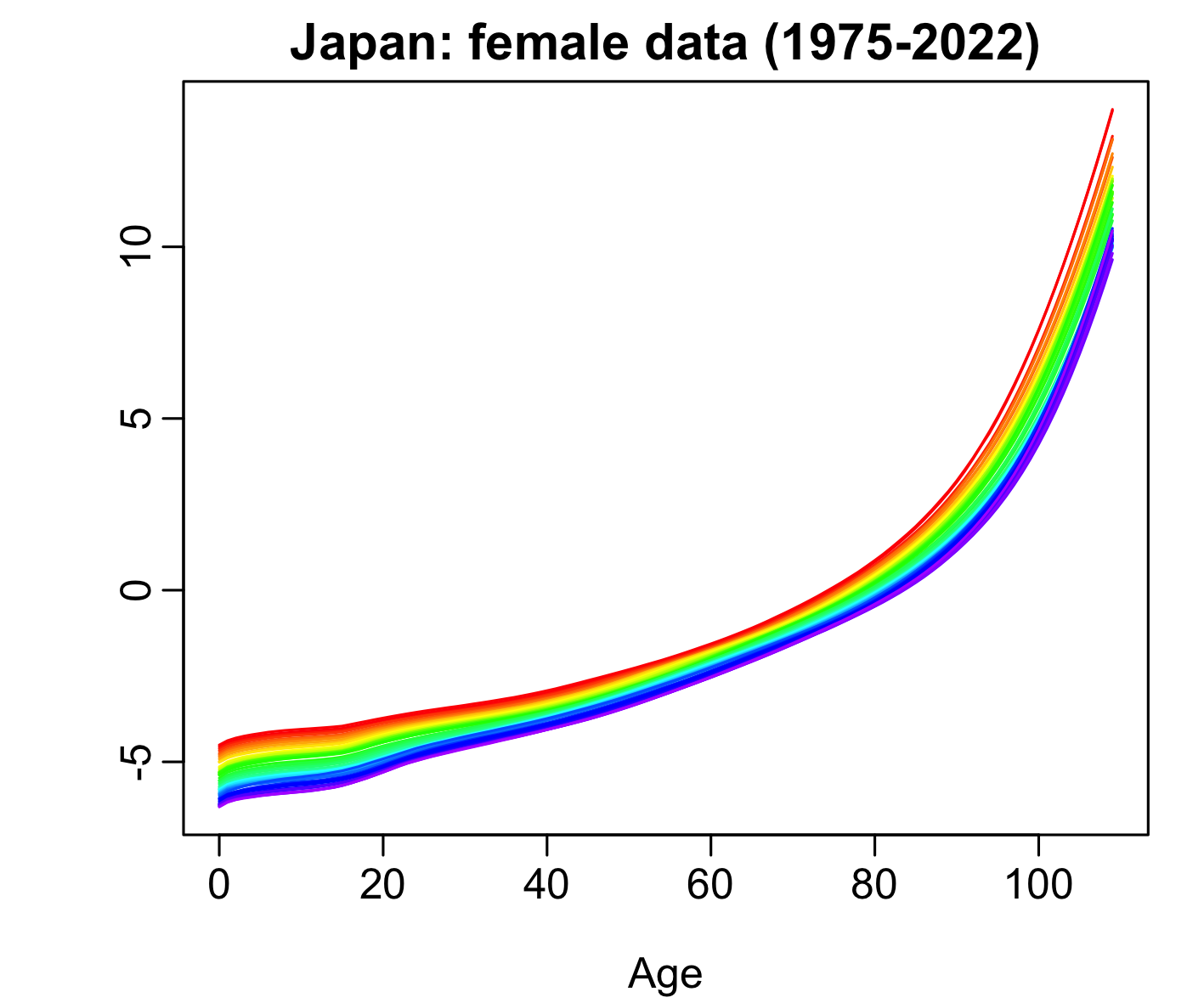}
\caption{\small{Empirical CDFs of the Japanese age- and sex-specific life-table death counts from 1975 to 2022.\label{fig:4}}}
\end{figure}
\item[3a)] Univariate functional time series (UFTS) method: With a set of unconstrained data $\bm{Z}^{s} = (\bm{Z}^{s}_{1},\dots, \bm{Z}^{s}_{n})^{\top}$ and $\bm{Z}^s_1 = (Z^{s}_{1,1}, \dots, Z^{s}_{n,111})$, we consider a functional time series forecasting technique \citep[see, e.g.,][]{KR17, HM24}. It is assumed that the observations $\bm{Z}^{s}$ are elements of the Hilbert space equipped with an inner product. Each function is a square-integrable function with a finite variance.

Assuming the existence of the mean and finite variance, its non-negative definite covariance function is given by
\begin{align*}
\text{Cov}^s(x,y) &:= \text{Cov}(\bm{Z}_{x}^s, \bm{Z}_{y}^s) \\
&=E(Z_x^s - \mu^s_x)(Z_y^s - \mu^s_y),
\end{align*}
where $\mu^s_x = \frac{1}{n}\sum^n_{t=1}Z_{t,x}^s$, $x$ or $y$ represents a continuum in a function support range $\mathcal{I}\in R$ and $R$ is the real line. Via Mercer's lemma, there exists a set of orthonormal sequence $\psi^s_{k,x}$ of continuous function and a non-increasing sequence $\lambda_k$ of positive numbers, such that
\begin{equation*}
\text{Cov}^{s}(x,y) = \sum^{\infty}_{\lambda=1}\lambda_k^s\psi_{k,x}^s\psi_{k,y}^s,
\end{equation*}
where $\lambda_k^s$ and $\psi_{k,x}^s$ denote the $k$\textsuperscript{th} eigenvalue and eigenfunction, respectively. The Karhunen-Lo\`{e}ve expansion of a stochastic process $Z^{s}_{t,x}$ can be expressed as
\begin{equation*}
Z^{s}_{t,x} = \sum^{K_s}_{k=1}\eta^{s}_{t,k}\psi^{s}_{k,x}+\varepsilon^{s}_{t,x},
\end{equation*}
where $\psi^{s}_{k,x}$ denotes the $k$\textsuperscript{th} orthonormal principal component for age $x$ and sex $s$, $\eta^{s}_{t,k}=\langle Z^{s}_{t,x},\psi^{s}_{k,x}\rangle$ is the estimated principal component score at time $t$, $\varepsilon^{s}_{t,x}$ denotes the model residual function for age~$x$ and sex~$s$ in year~$t$, and $K_s$ denotes the number of retained functional principal components. We consider an eigenvalue ratio (EVR) criterion of \cite{LRS20} to select the optimal value of $K_{s}$, as well as $K_{s}=6$ used in \cite{HBY13}, which is the default opinion in the demography package \citep{Hyndman23} in \Rlogo \ \citep{Team24}.
\item[3b)] Multivariate functional time series (MFTS) method: The female and male series are stacked in a vector $\bm{Z}_{t,\zeta} = (Z^{\text{F}}_{t,x}, Z^{\text{M}}_{t,x})\in R^2$ for a given age $x$. The number of ages in the stacked series is the sum of the number of ages in the two series. Before performing an eigendecomposition to $\bm{Z} = (\bm{Z}_{1,\zeta},\dots,\bm{Z}_{n,\zeta})^{\top}$, it is essential to centre each series by removing its mean and standard deviation. Applying eigendecomposition to $\bm{Z}$, it gives
\begin{equation*}
Z_{t,x}^{s} = \sum^{K}_{k=1}\eta_{t,k}^{s}\psi_{k,x}^{s} + \epsilon_{t,x}^{s},
\end{equation*}
where $\psi_{k,x}^{\text{F}}$ corresponds to the first 111 rows of the estimated principal components, while $\psi_{k,x}^{\text{M}}$ corresponds to the remaining 111 rows, and $K$ denotes the common number of principal components shared by both series. The value of $K$ can be determined by the EVR criterion or fixed at $K=6$.
\item[3c)] Multilevel functional time series (MLFTS) method: The joint modelling of the female and male series bears a strong resemblance to the two-way analysis of variance studied by \cite{Zhang13} and \cite{JSS24}. In the MLFTS method, the essential idea is to decompose each series into its mean function, a common trend across series $R^{(c)}$, and a series-specific residual trend $U^{s}$. It can be expressed as
\begin{equation*}
Z_{t,x}^{s} = \mu_{x}^s+R_{t,x}^{c}+U_{t,x}^{s},
\end{equation*}
where $\bm{R}_{x}^{c} = (R_{1,x}^{c},\dots, R_{n,x}^{c})^{\top}$ and $\bm{U}_{x}^{s}=(U_{1,x}^{s},\dots,U_{n,x}^{s})^{\top}$ can be modeled via two-stage principal component analysis. To model the common patterns effectively, identifying an auxiliary series that both series share is crucial. We suggest using a simple average of the series as a starting point for the first-stage functional principal component analysis. Then, each residual series is independently modelled by the second-stage functional principal component analysis.
\item[4)] Forecasting $Z^{s}_{n+h,x}$: By conditioning on the estimated functional principal components $\bm{\Psi}_x = (\psi_{1,x},\dots,\psi_{K,x})$ and observed data $\bm{Z}_x^s=(Z_{1,x}^s,\dots,Z_{n,x}^s)$, the $h$-step-ahead forecast of $Z_{n+h,x}^s$ can be obtained
\begin{equation*}
\widehat{Z}^{s}_{n+h|n,x} = \text{E}[Z^{s}_{n+h,x}|\bm{\Psi}_x,\bm{Z}_x] = \sum^{K}_{k=1}\widehat{\eta}^{s}_{n+h|n,k}\psi^{s}_{k,x},
\end{equation*}
where $\widehat{\eta}^{s}_{n+h|n,k}$ denotes the $h$-step-ahead univariate time-series forecast of the principal component scores. Among the univariate time-series methods, we consider the exponential smoothing method of \cite{HKO+08} for forecasting scores. In Figure~\ref{fig:5}, we present the 16 steps ahead forecasts of CDFs from 2023 to 2038 using the UFTS method.
\begin{figure}[!htb]
\centering
\includegraphics[width=8.5cm]{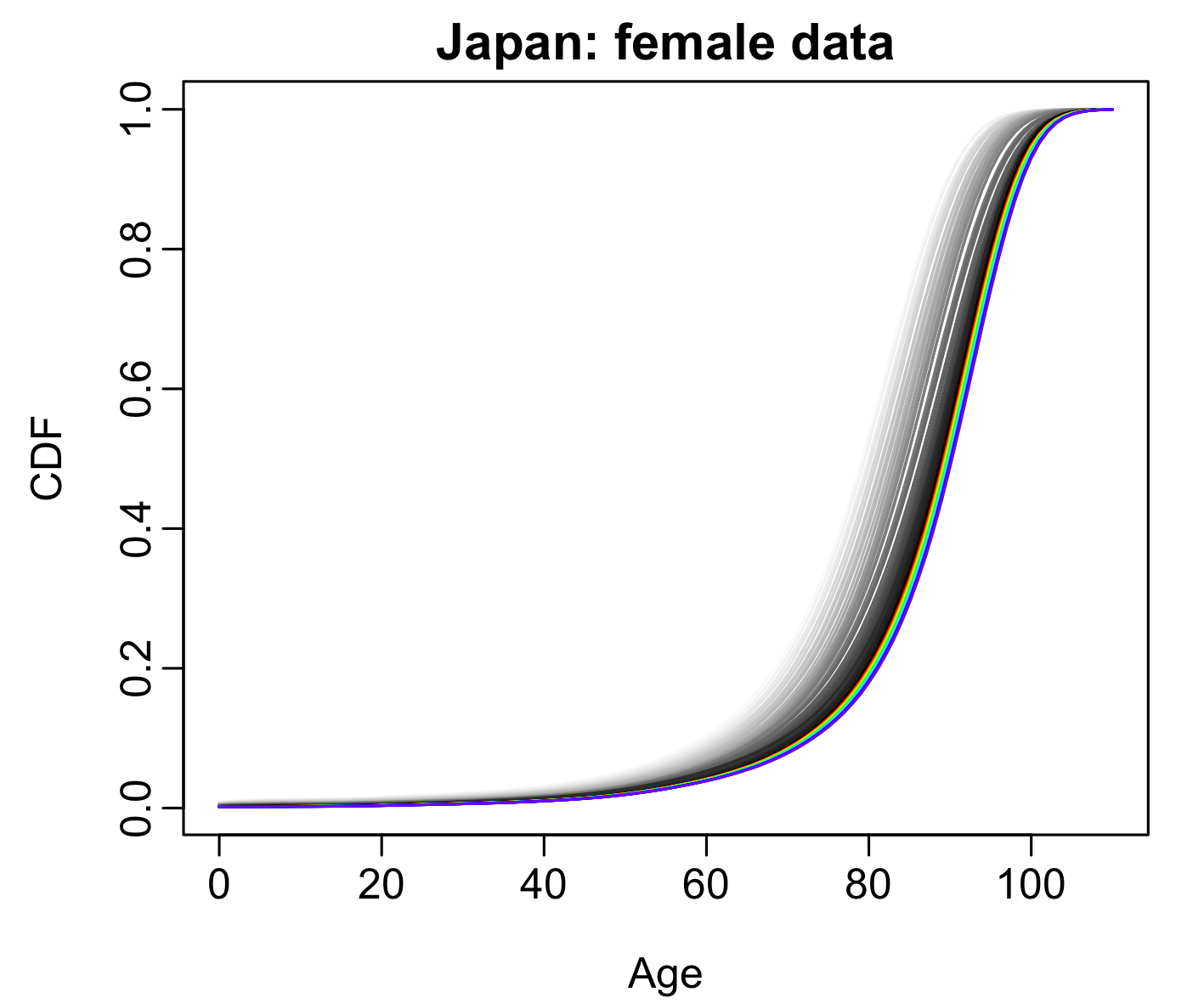}
\quad
\includegraphics[width=8.5cm]{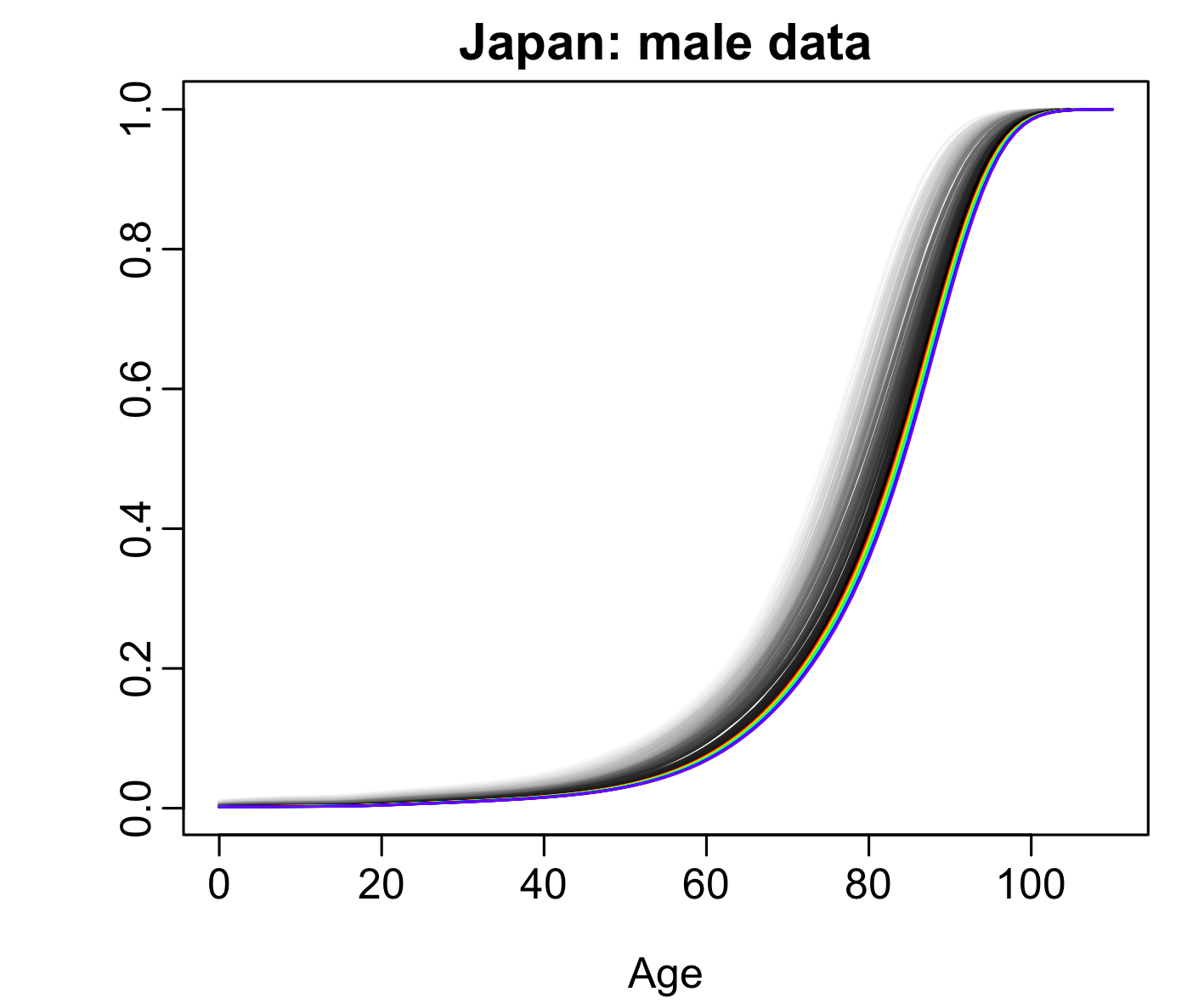}
\caption{\small{The 16 steps ahead forecasts of CDFs from 2023 to 2038 were obtained by the UFTS method. While the data from the distant past are shown in grey, the forecasts are shown in rainbow colours.}\label{fig:5}}
\end{figure}
\item[5)] By taking the inverse logit transformation, we obtain
\begin{equation*}
\widehat{D}^{s}_{n+h|n,x} = \frac{\exp^{\widehat{Z}^{s}_{n+h|n,x}}}{1+\exp^{\widehat{Z}^{s}_{n+h|n, x}}}
\end{equation*}
\item[6)] By taking the first-order differencing, we obtain
\begin{equation*}
\widehat{d}^{s}_{n+h|n,x} = \Delta_{i=1}^{x} \widehat{D}^{s}_{n+h|n,i},
\end{equation*}
where $\Delta$ represents the first-order differencing. In Figure~\ref{fig:6}, we display the 16 steps ahead forecasts of life-table death counts.
\begin{figure}[!htb]
\centering
\includegraphics[width=8.5cm]{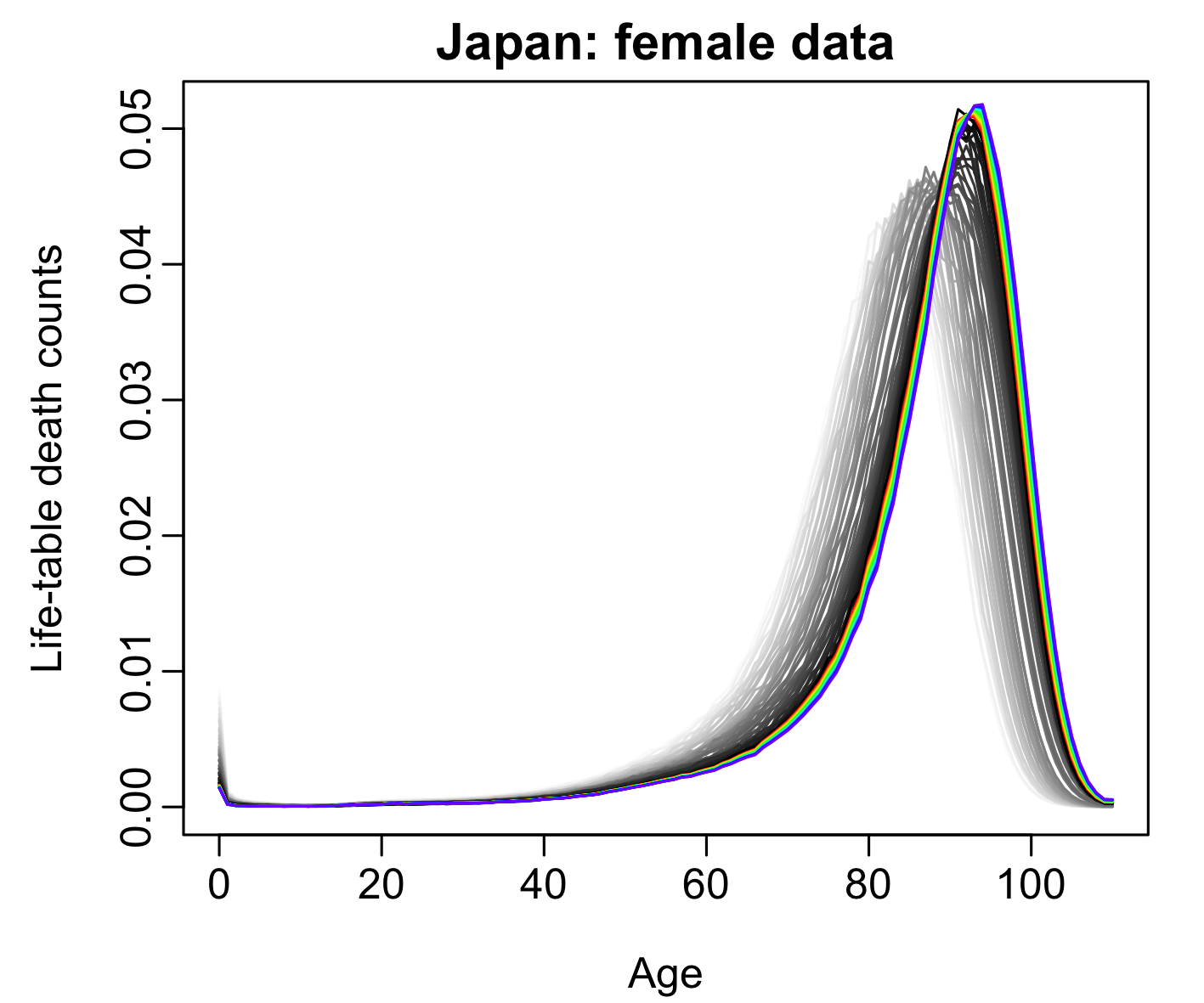}
\quad
\includegraphics[width=8.5cm]{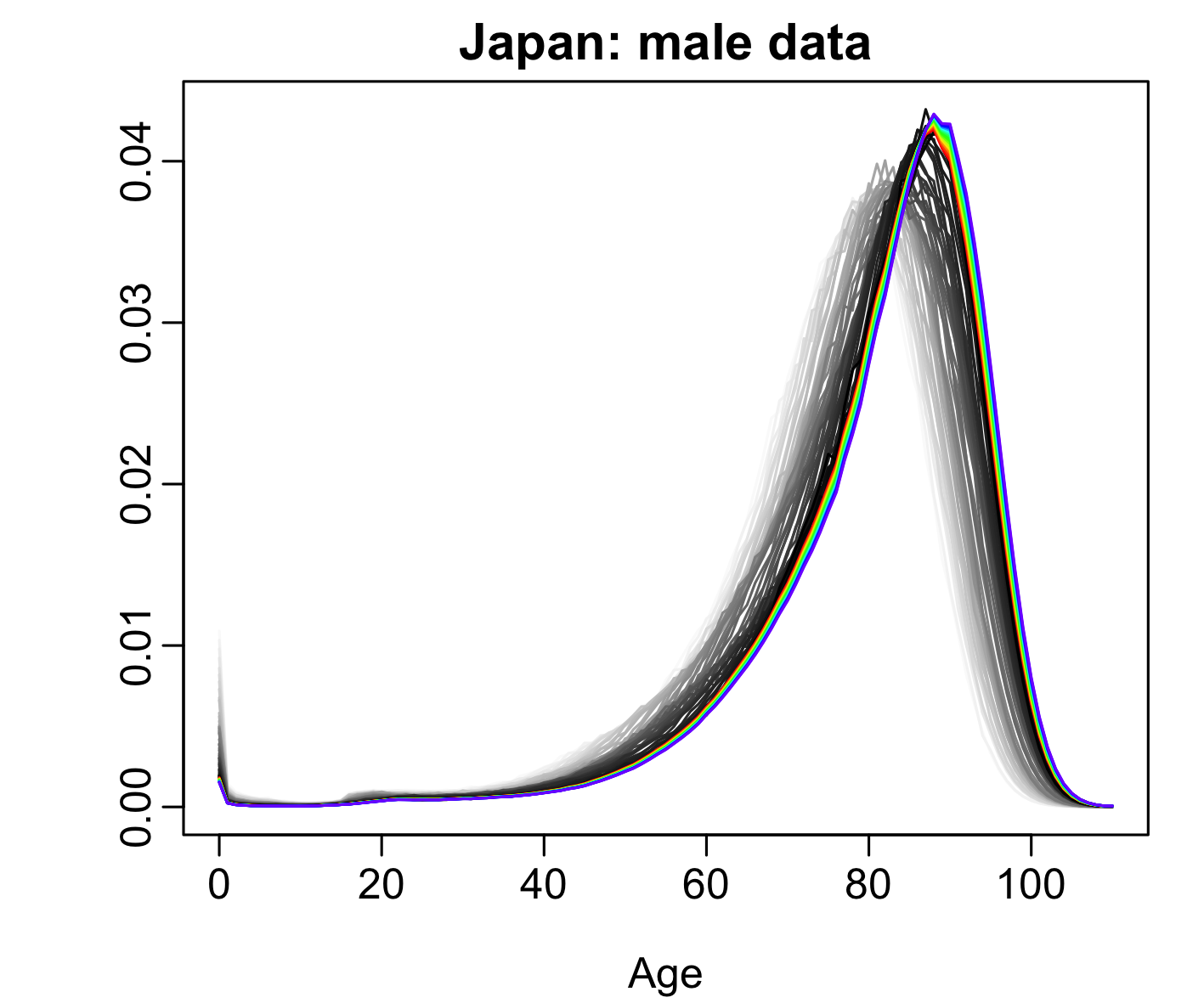}
\caption{\small{The 16 steps ahead forecasts of the life-table death counts from 2023 to 2038.}\label{fig:6}}
\end{figure}
\end{enumerate}

\subsection{Centered log-ratio transformation}\label{sec:3.2}

As a benchmark, the centred log-ratio transformation presents one of many possible ways to deal with the non-negativity constraint by transforming the raw data. The clr transformation \citep[see, e.g.,][]{Aitchison82, Aitchison86} is a mapping between the simplex to a hyperplane in Euclidean space. PCA can be applied directly to this hyperplane. The algorithm for implementing the CoDa method consists of the following steps:
\begin{enumerate}
\item[1)] Compute the geometric mean function, which can be estimated by a simple average
\begin{equation*}
\alpha_{n,x} = \exp\Big(\frac{1}{n}\sum^{n}_{t=1}\ln d_{t,x}\Big),
\end{equation*}
where $d_{t,x}>0$ denotes the age-specific life-table death counts. We compute the centred data by dividing each death count by its geometric mean
\begin{equation*}
s_{t,x} = \frac{d_{t,x}}{\alpha_{n,x}}.
\end{equation*}
\item[2)] By taking the nature logarithm, we obtain 
\begin{align*}
\beta_{t,x} &= \ln s_{t,x} \\
&= \ln d_{t,x} - \ln \alpha_{n,x} \\
&= \ln d_{t,x} - \frac{1}{n}\sum^{n}_{t=1}\ln d_{t,x}
\end{align*}
where $\bm{\beta}_x = (\beta_{1,x},\dots,\beta_{n,x})^{\top}$ can be viewed as an unconstrained functional time series.
\item[3)] Applying eigendecomposition to $\bm{\beta}_x$ gives
\begin{equation}
\beta_{t,x} = \sum^{K}_{k=1}\gamma_{t,k}\phi_{k,x} + \omega_{t,x}, \label{eq:clr}
\end{equation}
where $\phi_{k,x}$ is the $k$\textsuperscript{th} estimated principal component for age~$x$, $\gamma_{t,k} = \langle \beta_{t,x}, \phi_{k,x}\rangle$ is the $k$\textsuperscript{th} estimated principal component score at time $t$, and $\omega_{t,x}$ denotes the model residual function for age $x$ in year $t$. When $K=1$,~\eqref{eq:clr} mimics the spirit of \citeauthor{LC92}'s \citeyearpar{LC92} method. Within the CoDa framework, this method is known as the Lee-Carter-CoDa method.
\item[4)] Forecasting $\beta_{n+h,x}$: By conditioning on the estimated functional principal components $\bm{\Phi}_x = (\phi_{1,x},\dots,\phi_{K,x})$ and observed data $\bm{\beta}_x$, the $h$-step-ahead forecast of $\beta_{n+h,x}$ can be obtained
\begin{equation*}
\widehat{\beta}_{n+h|n,x} = \text{E}[\beta_{n+h,x}|\bm{\Phi}_x,\bm{\beta}_x] = \sum^K_{k=1}\widehat{\gamma}_{n+h|n,k}\phi_{k,x},
\end{equation*}
where $\widehat{\gamma}_{n+h|n,k}$ denotes the $h$-step-ahead univariate time-series forecast of the principal component scores.
\item[5)] Transform back to the compositional data. We take the inverse-centered log-ratio transformation given by
\begin{equation*}
\widehat{s}_{n+h|n,x} = \exp^{\widehat{\beta}_{n+h|n,x}}
\end{equation*}
\item[6)] We add back the geometric means to obtain the life-table death count forecasts $d_{n+h,x}$,
\begin{equation*}
\widehat{d}_{n+h|n,x} = \widehat{s}_{n+h|n,x}\times \alpha_{n,x},
\end{equation*}
where $\alpha_{n,x}$ denotes the simple mean given in Step~1).
\end{enumerate}

\section{Analysis of the Japanese life-table death counts}\label{sec:4}

\subsection{Expanding window}\label{sec:4.1}

An expanding window analysis of a time series model is commonly used to assess model and parameter stability over time, and prediction accuracy. The expanding window analysis determines the constancy of a model's parameter by computing parameter estimates and their forecasts over an expanding window through the sample \citep[for details][pp. 313-314]{ZW06}. Using the first 32 observations from 1975 to 2006 in the Japanese female and male age-specific life-table death counts, we produce one- to 16-step-ahead forecasts. Through an expanding-window approach, we estimate the parameters in the time series forecasting models using the first 33 observations from 1975 to 2007. Forecasts from the estimated models are then produced for one- to 15-step-ahead point and interval forecasts. We iterate this process by increasing the sample size by one year until reaching the end of the data period in 2022. This process produces 16 one-step-ahead forecasts, 15 two-step-ahead forecasts, $\dots$, and one 16-step-ahead forecast. We compare these forecasts with the holdout samples to determine the accuracy of the out-of-sample forecast.

\subsection{Point forecast evaluation}\label{sec:4.2}

We consider the expanding window scheme to assess the point forecast as described in \citet[][Chapter 9]{ZW06}. As the age distribution of death counts can be considered a probability density function, we consider density evaluation metrics. They include discrete Kullback-Leibler divergence \citep{KL51} and the Jensen-Shannon divergence \citep{Shannon48}.

The Kullback-Leibler divergence measures information loss by replacing an unknown density with an approximation. For two probability density functions, denoted by $d_{n+\xi}(u)$ and $\widehat{d}_{n+\xi|n}(u)$, the discrete Kullback-Leibler divergence is defined as
\begin{align*}
\text{KLD}(h) = \ & \text{D}_{\text{KL}}(d_{n+\xi,x}||\widehat{d}_{n+\xi|n,x})+\text{D}_{\text{KL}}(\widehat{d}_{n+\xi|n,x}||d_{n+\xi,x}) \\
		     = \ &\frac{1}{111\times (17-h)}\sum^{16}_{\xi=h}\sum^{111}_{x=1}d_{n+\xi,x}\cdot (\ln d_{n+\xi,x}-\ln \widehat{d}_{n+\xi|n,x}) + \\
			  &\frac{1}{111\times (17-h)}\sum^{16}_{\xi=h}\sum^{111}_{x=1}\widehat{d}_{n+\xi|n,x}\cdot (\ln \widehat{d}_{n+\xi|n,x} - \ln d_{n+\xi,x}),
\end{align*}
where $\xi$ denotes the forecasting period. The discrete Kullback-Leibler divergence is symmetric and non-negative.

An alternative is given by the Jensen-Shannon divergence, which is a symmetrised and smoothed version of the Kullback-Leibler divergence. It is defined by
\begin{equation*}
\text{JSD}(h) = \frac{1}{2}\text{D}_{\text{KL}}(d_{n+\xi,x}||\delta_{n+\xi,x})+\frac{1}{2}\text{D}_{\text{KL}}(\widehat{d}_{n+\xi|n,x}||\delta_{n+\xi,x}),
\end{equation*}
where $\delta_{n+\xi,x}$ measures a common quantity between $d_{n+\xi,x}$ and $\widehat{d}_{n+\xi,x}$. We consider the geometric mean $\delta_{n+\xi,x} = \sqrt{d_{n+\xi,x}\widehat{d}_{n+\xi|n,x}}$.

\subsection{Interval forecast evaluation}\label{sec:4.3}

For each year in the forecasting period, the $h$-step-ahead prediction intervals were computed at the $(1-\alpha)$ nominal coverage probabilities, where $\alpha$ denotes a level of significance. Let us denote $\widehat{d}^{\text{lb}}_{n+\xi|n,x}$ and $\widehat{d}^{\text{ub}}_{n+\xi|n,x}$ as the lower and upper bounds, respectively. We compute the empirical coverage probability, defined as the proportion of observations that fall into the calculated prediction intervals. It can be expressed as
\[
\text{ECP}_{\alpha}(h) = 1-\frac{\sum^{16}_{\xi=h}\sum^{111}_{x=1}\left[\mathds{1}(d_{n+\xi,x}<\widehat{d}^{\text{lb}}_{n+\xi |n,x})+\mathds{1}(d_{n+\xi,x}>\widehat{d}_{n+\xi |n,x}^{\text{ub}})\right]}{111\times (17-h)},
\]
where the denominator is the total number of observations in the forecasting period, and $\mathds{1}(\cdot)$ represents the binary indicator function. 

We compute the coverage probability difference (CPD), which is the absolute difference between the nominal and empirical coverage probabilities, and use this measure to evaluate the accuracy of the interval forecast \citep[see also][]{SB94, TSL07}. The CPD can be expressed as
\begin{equation*}
\text{CPD}_{\alpha}(h) = |\text{ECP}_{\alpha}(h) - (1-\alpha)|.
\end{equation*}
The smaller the value of CPD is, the more accurate the interval forecast one method produces.

Although the ECP and CPD are measures of interval forecast accuracy, neither consider the sharpness of the prediction intervals, i.e., the distance between the lower and upper bounds.  We also consider the interval score of \cite{GR07}, which is defined as
\begin{align*}
S_{\alpha}(\widehat{d}^{\text{lb}}_{n+\xi|n,x}, \widehat{d}^{\text{ub}}_{n+\xi|n,x}, d_{n+\xi,x}) = (\widehat{d}^{\text{ub}}_{n+\xi|n,x} - \widehat{d}^{\text{lb}}_{n+\xi|n,x})&+\frac{2}{\alpha}(\widehat{d}^{\text{lb}}_{n+\xi|n,x}-d_{n+\xi,x})\mathds{1}(d_{n+\xi,x}<\widehat{d}^{\text{lb}}_{n+\xi|n,x}) \\
&+  \frac{2}{\alpha}(d_{n+\xi,x} - \widehat{d}^{\text{ub}}_{n+\xi|n,x})\mathds{1}(d_{n+\xi,x} > \widehat{d}^{\text{ub}}_{n+\xi|n,x}).
\end{align*}
The interval score rewards a narrow prediction interval if and only if the true observation lies within the prediction interval. The optimal interval score is achieved when $d_{n+\xi,x}$ lies between $\widehat{d}^{\text{lb}}_{n+\xi|n,x}$ and $\widehat{d}^{\text{ub}}_{n+\xi|n,x}$ and the distance between $\widehat{d}^{\text{lb}}_{n+\xi|n,x}$ and $\widehat{d}^{\text{ub}}_{n+\xi|n,x}$ is minimal for age $x$.

For different ages and years in the forecasting period, the mean interval score is defined by
\begin{equation*}
\overline{S}_{\alpha}(h) = \frac{\sum^{16}_{\xi=h}\sum^{111}_{x=1}S_{\alpha}(\widehat{d}_{n+\xi|n,x}^{\text{lb}}, \widehat{d}_{n+\xi|n,x}^{\text{ub}}; d_{n+\xi,x})}{111\times (17-h)},
\end{equation*}
where $S_{\alpha}(\widehat{d}^{\text{lb}}_{n+\xi|n,x}, \widehat{d}^{\text{ub}}_{n+\xi|n,x}, d_{n+\xi,x})$ denotes the interval score at the $\xi$\textsuperscript{th} curve in the testing period.

\subsection{Comparisons of point forecast accuracy}\label{sec:4.4}

In Table~\ref{tab:1}, we present the one- to 16-step-ahead point forecast errors of the life-table death counts in Japan. The errors are measured by the symmetrised Kullback-Leibler Divergence (KLD) and Jensen-Shannon Divergence (JSD), which are commonly used for evaluating density forecasts. Since the errors were minuscule, we multiplied them by 100 to highlight the differences among the methods. Between the CDF and clr transformations, the CDF transformation coupled with the MLFTS method produces the smallest errors for almost all forecast horizons. Between the two methods of selecting the number of components, using $K=6$ proves advantageous, confirming the early finding of \cite{HBY13}. Additionally, between the female and male series, the latter yields smaller errors. For male data, the clr transformation performs better at longer horizons. Although that could be due to data randomness, the finding motivates us to consider a model-averaging approach by assigning weights to forecasts obtained from the two transformations.

\begin{center}
\tabcolsep 0.15in
\begin{longtable}{@{}llrrrrrrrr@{}}
\caption{\small{Comparison of the point forecast errors $(\times 100)$ of the age distribution of deaths between the CDF and clr transformations, when the number of components is fixed at $K=6$. Among the CDF transformation, we also consider the UFTS, MFTS and MLFTS methods. The smallest errors are shown in bold.}} \label{tab:1} \\
\toprule
	& & \multicolumn{2}{c}{UFTS} & \multicolumn{2}{c}{MFTS} & \multicolumn{2}{c}{MLFTS} & \multicolumn{2}{c}{clr}  \\
Sex & $h$ & KLD & JSD & KLD & JSD & KLD & JSD & KLD & JSD \\ 
	\midrule
\endfirsthead
\toprule
	& & \multicolumn{2}{c}{UFTS} & \multicolumn{2}{c}{MFTS} & \multicolumn{2}{c}{MLFTS} & \multicolumn{2}{c}{clr} \\
Sex &  $h$ & KLD & JSD & KLD & JSD & KLD & JSD & KLD & JSD \\ 
		\midrule
\endhead	
\hline \multicolumn{10}{r}{{Continued on next page}} \\
\endfoot
\endlastfoot
F 	& 1 & \textBF{0.0762} & \textBF{0.0193} & 0.1010 & 0.0255 & 0.0812 & 0.0206  & 0.0992 & 0.0252 \\ 
  	& 2 & 0.1313 & 0.0334 & 0.1617 & 0.0408 & \textBF{0.1175} &  \textBF{0.0300} & 0.1629 & 0.0414 \\ 
  	& 3 & 0.1994 & 0.0506 & 0.2129 & 0.0530 & \textBF{0.1492} &  \textBF{0.0380} & 0.2342 & 0.0593 \\ 
  	& 4 & 0.3039 & 0.0774 & 0.2997 & 0.0746 & \textBF{0.2105} &  \textBF{0.0539} & 0.3373 & 0.0857 \\ 
	& 5 & 0.4225 & 0.1078 & 0.4010 & 0.0991 & \textBF{0.2769} &  \textBF{0.0709} & 0.4466 & 0.1136 \\ 
  	& 6 & 0.5365 & 0.1370 & 0.4911 & 0.1203 & \textBF{0.3245} &  \textBF{0.0832} & 0.5582 & 0.1420 \\ 
  	& 7 & 0.6890 & 0.1749 & 0.5649 & 0.1369 & \textBF{0.3702} &  \textBF{0.0945} & 0.7334 & 0.1840 \\ 
 	& 8 & 0.9445 & 0.2375 & 0.5236 & 0.1297 & \textBF{0.3956} &  \textBF{0.1021} & 0.9441 & 0.2344 \\ 
  	& 9 & 1.1893 & 0.2994 & 0.5401 & 0.1376 & \textBF{0.4295} &  \textBF{0.1126} & 1.0786 & 0.2672 \\ 
  	& 10 & 1.2043 & 0.3103 & 0.5681 & 0.1499 & \textBF{0.4729} &  \textBF{0.1251} & 1.0099 & 0.2562 \\ 
 	& 11 & 1.0678 & 0.2842 & \textBF{0.5243} &  \textBF{0.1409} & 0.5909 & 0.1578 & 0.7761 & 0.2050 \\ 
  	& 12 & 1.2257 & 0.3368 & \textBF{0.7628} &  \textBF{0.2056} & 0.7742 & 0.2088 & 1.0570 & 0.2827 \\ 
  	& 13 & 1.7173 & 0.4787 & 1.0650 & 0.2924 & \textBF{0.9434} &  \textBF{0.2570} & 1.3935 & 0.3759 \\ 
  	& 14 & 2.0706 & 0.5826 & 1.3167 & 0.3667 & \textBF{1.0168} &  \textBF{0.2760} & 1.7221 & 0.4647 \\ 
  	& 15 & 3.0846 & 0.8887 & 1.6464 & 0.4619 & \textBF{1.6175} &  \textBF{0.4472} & 2.8137 & 0.7742 \\ 
  	& 16 & 4.0159 & 1.1744 & \textBF{2.3856} &  \textBF{0.6839} & 3.3945 & 0.9802 & 4.0931 & 1.1474 \\ 
\cmidrule{2-10}
  	& Mean & 1.1799 & 0.3246 & 0.7228 & 0.1949 & \textBF{0.6978} &  \textBF{0.1911} & 1.0912 & 0.2912 \\
\midrule
M 	& 1 & 0.0882 & 0.0222 & 0.0894 & 0.0225 & \textBF{0.0799} & \textBF{0.0202} & 0.0830 & 0.0209 \\ 
  	& 2 & 0.1212 & 0.0305 & 0.1346 & 0.0337 & \textBF{0.1116} & \textBF{0.0282} & 0.1270 & 0.0319 \\ 
  	& 3 & 0.1606 & 0.0403 & 0.1863 & 0.0463 & \textBF{0.1399} & \textBF{0.0352} & 0.1863 & 0.0465 \\ 
  	& 4 & 0.2093 & 0.0524 & 0.2575 & 0.0639 & \textBF{0.1889} & \textBF{0.0475} & 0.2652 & 0.0659 \\ 
  	& 5 & 0.2706 & 0.0677 & 0.3378 & 0.0834 & \textBF{0.2404} & \textBF{0.0605} & 0.3546 & 0.0880 \\ 
  	& 6 & 0.3500 & 0.0875 & 0.4231 & 0.1038 & \textBF{0.2931} & \textBF{0.0736} & 0.4575 & 0.1131 \\ 
  	& 7 & 0.4321 & 0.1075 & 0.4988 & 0.1215 & \textBF{0.3495} & \textBF{0.0873} & 0.5953 & 0.1459 \\ 
  	& 8 & 0.5309 & 0.1320 & 0.5326 & 0.1309 & \textBF{0.4075} & \textBF{0.1018} & 0.7716 & 0.1866 \\ 
  	& 9 & 0.6153 & 0.1538 & 0.5494 & 0.1354 & \textBF{0.4164} & \textBF{0.1046} & 0.9298 & 0.2233 \\ 
  	& 10 & 0.6590 & 0.1659 & 0.5204 & 0.1296 & \textBF{0.3345} & \textBF{0.0857} & 0.9376 & 0.2286 \\ 
  	& 11 & 0.6095 & 0.1550 & 0.4766 & 0.1209 & \textBF{0.3148} & \textBF{0.0814} & 0.6860 & 0.1706 \\ 
  	& 12 & 0.5926 & 0.1510 & 0.4342 & 0.1109 & \textBF{0.3261} & \textBF{0.0846} & 0.3517 & 0.0886 \\ 
  	& 13 & 0.6250 & 0.1599 & 0.4773 & 0.1230 & 0.3630 & 0.0952 & \textBF{0.3436} & \textBF{0.0866} \\ 
  	& 14 & 0.6539 & 0.1677 & 0.4701 & 0.1209 & 0.3914 & 0.1032 & \textBF{0.3110} & \textBF{0.0788} \\ 
  	& 15 & 0.6204 & 0.1605 & 0.5733 & 0.1506 & 0.5622 & 0.1523 & \textBF{0.2314} & \textBF{0.0594} \\ 
  	& 16 & 0.5630 & 0.1471 & 0.6282 & 0.1666 & 0.8016 & 0.2191 & \textBF{0.2417} & \textBF{0.0641} \\ 
\cmidrule{2-10}
 	& Mean & 0.4439 & 0.1126 & 0.4119 & 0.1040 & \textBF{0.3325} & \textBF{0.0863} & 0.4296 & 0.1062 \\
\bottomrule
\end{longtable}
\end{center}   

\vspace{-.5in}

In Table~\ref{tab:e0_1}, we display a comparison of the point forecast errors of the life expectancy using a terminal age of 111. Based on the forecasted age distribution of deaths from the CDF and clr transformations, we compute their corresponding age-specific life expectancies and compare these values with the holdout data. To measure the forecast accuracy, we consider the root mean squared forecast error (RMSFE) and mean absolute forecast error (MAFE). For forecasting age-specific life expectancy, the CDF transformation coupled with the MLFTS method is recommended.

\begin{center}
\tabcolsep 0.125in
\begin{longtable}{@{}llrrrrrrrr@{}}
\caption{\small{Comparison of the point forecast errors of the life expectancy between the CDF and clr transformations, when the number of components is fixed at $K=6$.}} \label{tab:e0_1} \\
\toprule
	& & \multicolumn{2}{c}{UFTS} & \multicolumn{2}{c}{MFTS} & \multicolumn{2}{c}{MLFTS} & \multicolumn{2}{c}{clr}  \\
Sex & $h$ & RMSFE & MAFE & RMSFE & MAFE & RMSFE & MAFE & RMSFE & MAFE \\ 
	\midrule
\endfirsthead
\toprule
	& & \multicolumn{2}{c}{UFTS} & \multicolumn{2}{c}{MFTS} & \multicolumn{2}{c}{MLFTS} & \multicolumn{2}{c}{clr} \\
Sex &  $h$ & RMSFE & MAFE & RMSFE & MAFE & RMSFE & MAFE & RMSFE & MAFE \\ 
		\midrule
\endhead	
\hline \multicolumn{10}{r}{{Continued on next page}} \\
\endfoot
\endlastfoot
F & 1 & 0.2358 & 0.1756 & 0.2555 & 0.1920 & \textBF{0.2142} & \textBF{0.1458} & 0.2861 & 0.1953 \\ 
   & 2 & 0.3061 & 0.2186 & 0.3127 & 0.2485 & \textBF{0.2451} & \textBF{0.1770} & 0.3732 & 0.2715 \\ 
   & 3 & 0.3776 & 0.2724 & 0.3323 & 0.2885 & \textBF{0.2566} & \textBF{0.1987} & 0.4529 & 0.3458 \\ 
   & 4 & 0.4738 & 0.3446 & 0.3790 & 0.3395 & \textBF{0.3041} & \textBF{0.2502} & 0.5512 & 0.4318 \\ 
   & 5 & 0.5598 & 0.4129 & 0.4247 & 0.3708 & \textBF{0.3461} & \textBF{0.2770} & 0.6354 & 0.5152 \\ 
   & 6 & 0.6224 & 0.4559 & 0.4467 & 0.3726 & \textBF{0.3340} & \textBF{0.2716} & 0.7042 & 0.5829 \\ 
   & 7 & 0.7139 & 0.5445 & 0.4581 & 0.3654 & \textBF{0.3442} & \textBF{0.2786} & 0.8125 & 0.6685 \\ 
   & 8 & 0.8542 & 0.6361 & 0.4172 & 0.3565 & \textBF{0.3764} & \textBF{0.2907} & 0.9254 & 0.7777 \\ 
   & 9 & 0.9556 & 0.7346 & 0.4002 & 0.3355 & \textBF{0.3956} & \textBF{0.3159} & 0.9673 & 0.8301 \\ 
   & 10 & 0.9274 & 0.7281 & 0.4143 & 0.3663 & \textBF{0.3068} & \textBF{0.2382} & 0.9263 & 0.8147 \\ 
   & 11 & 0.7744 & 0.6485 & 0.3338 & 0.2892 & \textBF{0.3109} & \textBF{0.2383} & 0.7865 & 0.7195 \\ 
   & 12 & 0.7065 & 0.6729 & 0.4767 & 0.4390 & \textBF{0.4646} & \textBF{0.3803} & 0.9765 & 0.8982 \\ 
   & 13 & 0.8900 & 0.8316 & 0.5998 & 0.5229 & \textBF{0.5241} & \textBF{0.4265} & 1.1285 & 1.0424 \\ 
   & 14 & 1.0003 & 0.9321 & 0.6986 & 0.5830 & \textBF{0.5202} & \textBF{0.4625} & 1.2804 & 1.1860 \\ 
   & 15 & 1.2642 & 1.1776 & 0.7784 & \textBF{0.7053} & \textBF{0.7311} & 0.7071 & 1.6265 & 1.5007 \\ 
   & 16 & 1.4548 & 1.3989 & \textBF{0.9476} & \textBF{0.9200} & 1.2572 & 1.2121 & 1.9369 & 1.8094 \\ 
  \cmidrule{2-10}
   & Mean & 0.7573 & 0.6366 & 0.4797 & 0.4184 & \textBF{0.4332} & \textBF{0.3669} & 0.8981 & 0.7869 \\ 
\midrule
M & 1 & 0.2373 & 0.1898 & 0.2303 & 0.1726 & \textBF{0.2002} & \textBF{0.1331} & 0.2476 & 0.1918 \\ 
    & 2 & 0.2788 & 0.2259 & 0.2738 & 0.2140 & \textBF{0.2343} & \textBF{0.1625} & 0.3250 & 0.2475 \\ 
    & 3 & 0.3236 & 0.2649 & 0.3104 & 0.2506 & \textBF{0.2720} & \textBF{0.1960} & 0.4160 & 0.3042 \\ 
    & 4 & 0.3754 & 0.3038 & 0.3615 & 0.3010 & \textBF{0.3219} & \textBF{0.2311} & 0.5160 & 0.3647 \\ 
    & 5 & 0.4408 & 0.3529 & 0.4128 & 0.3444 & \textBF{0.3699} & \textBF{0.2762} & 0.6170 & 0.4460 \\ 
    & 6 & 0.5136 & 0.4187 & 0.4738 & 0.4105 & \textBF{0.4122} & \textBF{0.3181} & 0.7188 & 0.5161 \\ 
    & 7 & 0.5854 & 0.4713 & 0.5268 & 0.4606 & \textBF{0.4651} & \textBF{0.3713} & 0.8369 & 0.5934 \\ 
    & 8 & 0.6681 & 0.5376 & 0.5696 & 0.4904 & \textBF{0.5314} & \textBF{0.4379} & 0.9646 & 0.6740 \\ 
    & 9 & 0.7655 & 0.6147 & 0.6034 & 0.5063 & \textBF{0.5573} & \textBF{0.4675} & 1.0710 & 0.7540 \\ 
    & 10 & 0.8418 & 0.7122 & 0.6248 & 0.5235 & \textBF{0.4731} & \textBF{0.3893} & 1.0932 & 0.8219 \\ 
    & 11 & 0.8411 & 0.7125 & 0.6244 & 0.5209 & \textBF{0.4381} & \textBF{0.3526} & 0.9364 & 0.7041 \\ 
    & 12 & 0.8274 & 0.6877 & 0.6399 & 0.5116 & \textBF{0.3994} & \textBF{0.3322} & 0.6205 & 0.4960 \\ 
    & 13 & 0.8338 & 0.6825 & 0.6620 & 0.5316 & \textBF{0.3853} & \textBF{0.3115} & 0.6218 & 0.5038 \\ 
    & 14 & 0.8406 & 0.6753 & 0.6649 & 0.5246 & \textBF{0.3871} & \textBF{0.3180} & 0.5815 & 0.4670 \\ 
    & 15 & 0.7650 & 0.6142 & 0.6044 & 0.4757 & \textBF{0.3640} & \textBF{0.3010} & 0.4227 & 0.3554 \\ 
    & 16 & 0.5948 & 0.5013 & 0.4569 & 0.4068 & 0.3638 & 0.2777 & \textBF{0.1855} & \textBF{0.1570} \\ 
  \cmidrule{2-10}
   & Mean & 0.6083 & 0.4978 & 0.5025 & 0.4153 & \textBF{0.3859} & \textBF{0.3048} & 0.6359 & 0.4748 \\ 
\bottomrule
\end{longtable}
\end{center}   

\vspace{-.5in}

In Appendix~\ref{sec:appendix_A}, we present the comparisons of point forecast accuracy of the age distribution of deaths and life expectancy obtained from the CDF and clr transformations when the number of components is selected by the EVR criterion.

\subsection{Comparisons of interval forecast accuracy}\label{sec:4.5}

The interval forecast errors are measured by the mean interval score and coverage probability deviation (CPD). In Table~\ref{tab:2}, we present the one- to 16-step-ahead interval forecast errors of the life-table death counts in Japan under a nominal coverage probability of 80\% when the number of components is fixed at $K=6$. Between the CDF and clr transformations, the CDF transformation coupled with the MLFTS method produces the smallest interval scores for the shorter forecast horizon. Conversely, for the female series, the clr transformation results in the smallest errors over a relatively longer forecast horizon. Regarding the CPD, the CDF transformation with the MLFTS method is advantageous for shorter forecast horizons. For relatively longer forecast horizons, the CDF transformation using the univariate functional time series method is recommended. Additionally, between the female and male series, the latter yields smaller errors.

\begin{center}
\tabcolsep 0.082in
\renewcommand*{\arraystretch}{0.99}
\begin{longtable}{@{}llrrrrrrrrrr@{}}
\caption{\small{Comparison of the interval forecast errors between the CDF and clr transformations at the 80\% nominal coverage probability.}} \label{tab:2} \\
\toprule
	& & \multicolumn{2}{c}{UFTS} & \multicolumn{2}{c}{MFTS} & \multicolumn{2}{c}{MLFTS} & \multicolumn{2}{c}{clr (EVR)} & \multicolumn{2}{c}{clr ($K=6$)} \\
Sex & $h$ & score & CPD & score & CPD & score & CPD & score & CPD & score & CPD \\ 
	\midrule
\endfirsthead
\toprule
	& & \multicolumn{2}{c}{UFTS} & \multicolumn{2}{c}{MFTS} & \multicolumn{2}{c}{MLFTS} & \multicolumn{2}{c}{clr (EVR)} & \multicolumn{2}{c}{clr ($K=6$)} \\
Sex & $h$ & score & CPD & score & CPD & score & CPD & score & CPD & score & CPD \\ 
	\midrule
\endhead
\hline \multicolumn{12}{r}{{Continued on next page}} \\
\endfoot
\endlastfoot
F 	& 1 & 0.0016 & 0.0761 & 0.0019 & 0.1499 & \textBF{0.0013} & 0.0218 & 0.0017 & 0.0153 & 0.0015 & \textBF{0.0080} \\ 
  	& 2 & 0.0018 & 0.0787 & 0.0021 & 0.1303 & \textBF{0.0016} & 0.0174 & 0.0020 & \textBF{0.0006} & 0.0018 & 0.0186 \\ 
  	& 3 & 0.0022 & 0.0662 & 0.0023 & 0.1402 & \textBF{0.0019} & \textBF{0.0143} & 0.0024 & 0.0385 & 0.0022 & 0.0314 \\ 
  	& 4 & 0.0024 & 0.0517 & 0.0026 & 0.1279 & \textBF{0.0023} & \textBF{0.0162} & 0.0027 & 0.0219 & 0.0024 & 0.0344 \\ 
  	& 5 & 0.0029 & 0.0416 & 0.0030 & 0.1159 & \textBF{0.0028} & \textBF{0.0335} & 0.0031 & 0.0544 & 0.0029 & 0.0634 \\ 
  	& 6 & 0.0032 & 0.0583 & 0.0035 & 0.1230 & \textBF{0.0031} & \textBF{0.0051} & 0.0034 & 0.1181 & 0.0031 & 0.0903 \\ 
  	& 7 & 0.0036 & 0.0505 & 0.0038 & 0.1198 & 0.0035 & \textBF{0.0018} & 0.0036 & 0.1162 & \textBF{0.0035} & 0.1126 \\ 
  	& 8 & 0.0039 & 0.0368 & 0.0042 & 0.1279 & 0.0039 & \textBF{0.0248} & 0.0042 & 0.1299 & \textBF{0.0038} & 0.1229 \\ 
  	& 9 & 0.0041 & 0.0255 & 0.0047 & 0.1178 & 0.0043 & \textBF{0.0108} & 0.0044 & 0.1572 & \textBF{0.0040} & 0.1347 \\ 
  	& 10 & 0.0045 & 0.0417 & 0.0052 & 0.1228 & 0.0046 & \textBF{0.0391} & 0.0049 & 0.1562 & \textBF{0.0045} & 0.1447 \\ 
  	& 11 & 0.0048 & 0.0604 & 0.0059 & 0.1204 & 0.0049 & \textBF{0.0423} & 0.0052 & 0.1655 & \textBF{0.0047} & 0.1384 \\ 
  	& 12 & 0.0057 & 0.1009 & 0.0066 & 0.1495 & 0.0058 & \textBF{0.0793} & 0.0059 & 0.1586 & \textBF{0.0056} & 0.1351 \\ 
  	& 13 & 0.0067 & 0.1009 & 0.0074 & 0.1459 & 0.0069 & \textBF{0.0806} & 0.0070 & 0.1662 & \textBF{0.0067} & 0.1279 \\ 
  	& 14 & 0.0076 & 0.1009 & 0.0083 & 0.1369 & \textBF{0.0073} & \textBF{0.0589} & 0.0074 & 0.1610 & 0.0075 & 0.1159 \\ 
  	& 15 & 0.0085 & 0.1009 & 0.0094 & 0.1414 & 0.0089 & \textBF{0.0063} & \textBF{0.0083} & 0.1640 & 0.0086 & 0.1279 \\ 
  	& 16 & 0.0098 & 0.1009 & 0.0110 & 0.1459 & 0.0109 & \textBF{0.0432} & \textBF{0.0097} & 0.1550 & 0.0099 & 0.1279 \\ 
\cmidrule{2-12}
  	& Mean & 0.0046 & 0.0682 & 0.0051 & 0.1322 & 0.0046 & \textBF{0.0310} & 0.0047 & 0.1112 & \textBF{0.0045} & 0.0959 \\ 
\midrule
M 	& 1 & 0.0016 & 0.1459 & 0.0019 & 0.1662 & \textBF{0.0012} & \textBF{0.0078} & 0.0017 & 0.0367 & 0.0014 & 0.0654 \\ 
	& 2 & 0.0018 & 0.1345 & 0.0020 & 0.1604 & \textBF{0.0014} & \textBF{0.0018} & 0.0019 & 0.0505 & 0.0016 & 0.0775 \\ 
  	& 3 & 0.0021 & 0.1421 & 0.0021 & 0.1505 & \textBF{0.0016} & \textBF{0.0076} & 0.0023 & 0.0346 & 0.0019 & 0.0726 \\ 
  	& 4 & 0.0024 & 0.1362 & 0.0022 & 0.1612 & \textBF{0.0019} & \textBF{0.0141} & 0.0026 & 0.0420 & 0.0023 & 0.0884 \\ 
  	& 5 & 0.0027 & 0.1354 & 0.0024 & 0.1542 & \textBF{0.0023} & \textBF{0.0087} & 0.0029 & 0.0551 & 0.0025 & 0.0649 \\ 
  	& 6 & 0.0030 & 0.1181 & 0.0027 & 0.1443 & \textBF{0.0026} & \textBF{0.0075} & 0.0033 & 0.0452 & 0.0028 & 0.0739 \\ 
  	& 7 & 0.0033 & 0.0919 & \textBF{0.0029} & 0.1171 & 0.0030 & \textBF{0.0063} & 0.0036 & 0.0252 & 0.0031 & 0.0622 \\ 
  	& 8 & 0.0036 & 0.0709 & \textBF{0.0032} & 0.1149 & 0.0032 & 0.0268 & 0.0040 & \textBF{0.0238} & 0.0034 & 0.0569 \\ 
  	& 9 & 0.0037 & 0.0412 & 0.0034 & 0.0919 & \textBF{0.0033} & 0.0333 & 0.0042 & \textBF{0.0232} & 0.0034 & 0.0311 \\ 
  	& 10 & 0.0039 & \textBF{0.0211} & 0.0037 & 0.0829 & 0.0033 & 0.0842 & 0.0037 & 0.0520 & \textBF{0.0031} & 0.0327 \\ 
  	& 11 & 0.0040 & \textBF{0.0012} & 0.0041 & 0.0784 & 0.0035 & 0.0784 & 0.0034 & 0.1024 & \textBF{0.0032} & 0.0814 \\ 
  	& 12 & 0.0043 & \textBF{0.0234} & 0.0044 & 0.0667 & 0.0039 & 0.0973 & 0.0040 & 0.1820 & \textBF{0.0036} & 0.1459 \\ 
  	& 13 & 0.0048 & \textBF{0.0477} & 0.0048 & 0.0514 & 0.0044 & 0.1189 & 0.0043 & 0.1865 & \textBF{0.0038} & 0.1212 \\ 
  	& 14 & 0.0050 & \textBF{0.0372} & 0.0050 & 0.0859 & 0.0045 & 0.1069 & 0.0042 & 0.1880 & \textBF{0.0040} & 0.1489 \\ 
  	& 15 & 0.0053 & \textBF{0.0207} & 0.0057 & 0.1054 & 0.0050 & 0.1189 & 0.0047 & 0.1910 & \textBF{0.0046} & 0.1595 \\ 
  	& 16 & 0.0058 & \textBF{0.0072} & 0.0062 & 0.1550 & 0.0056 & 0.1459 & \textBF{0.0053} & 0.1820 & 0.0054 & 0.1640 \\ 
\cmidrule{2-12}
	&  Mean & 0.0036 & 0.0734 & 0.0035 & 0.1179 & 0.0032 & \textBF{0.0540} & 0.0035 & 0.0888 & \textBF{0.0031} & 0.0904 \\ 
\bottomrule
\end{longtable}
\end{center}   

\vspace{-.5in}

When the number of components is selected by the EVR criterion, the additional results are presented in Appendix~\ref{sec:appendix_B}. In Appendix~\ref{sec:appendix_C}, we also display the one- to 16-step-ahead interval forecast errors under a nominal coverage probability of 95\%.

\section{Application to a single-premium temporary immediate annuity}\label{sec:5}

An important application of mortality forecasts for those individuals at ages over 60 is in the pension and superannuation industries, whose profitability and solvency crucially depend on accurate forecasts to appropriately hedge longevity risks. When an individual retires, an optimal way of guaranteeing one individual's financial income in retirement is to purchase an annuity \citep{Yaari65}. An annuity is a financial contract offered by insurers that guarantees a steady stream of income payments for either a temporary or lifetime of the annuitants in exchange for an initial premium fee.

We study temporary annuities, which have gained popularity in developed countries (e.g., Australia and the USA), because lifetime immediate annuities, where rates are locked in for life, have been shown to deliver poor value for money \citep[see, e.g.,][Chapter 6]{CT08}. These temporary annuities pay a pre-determined and guaranteed income level higher than the income provided by a lifetime annuity for a similar premium. In combination with deferred annuities, fixed-term annuities offer an attractive alternative to lifetime immediate annuities.

\begin{figure}[!htb]
\centering
\includegraphics[width=8.6cm]{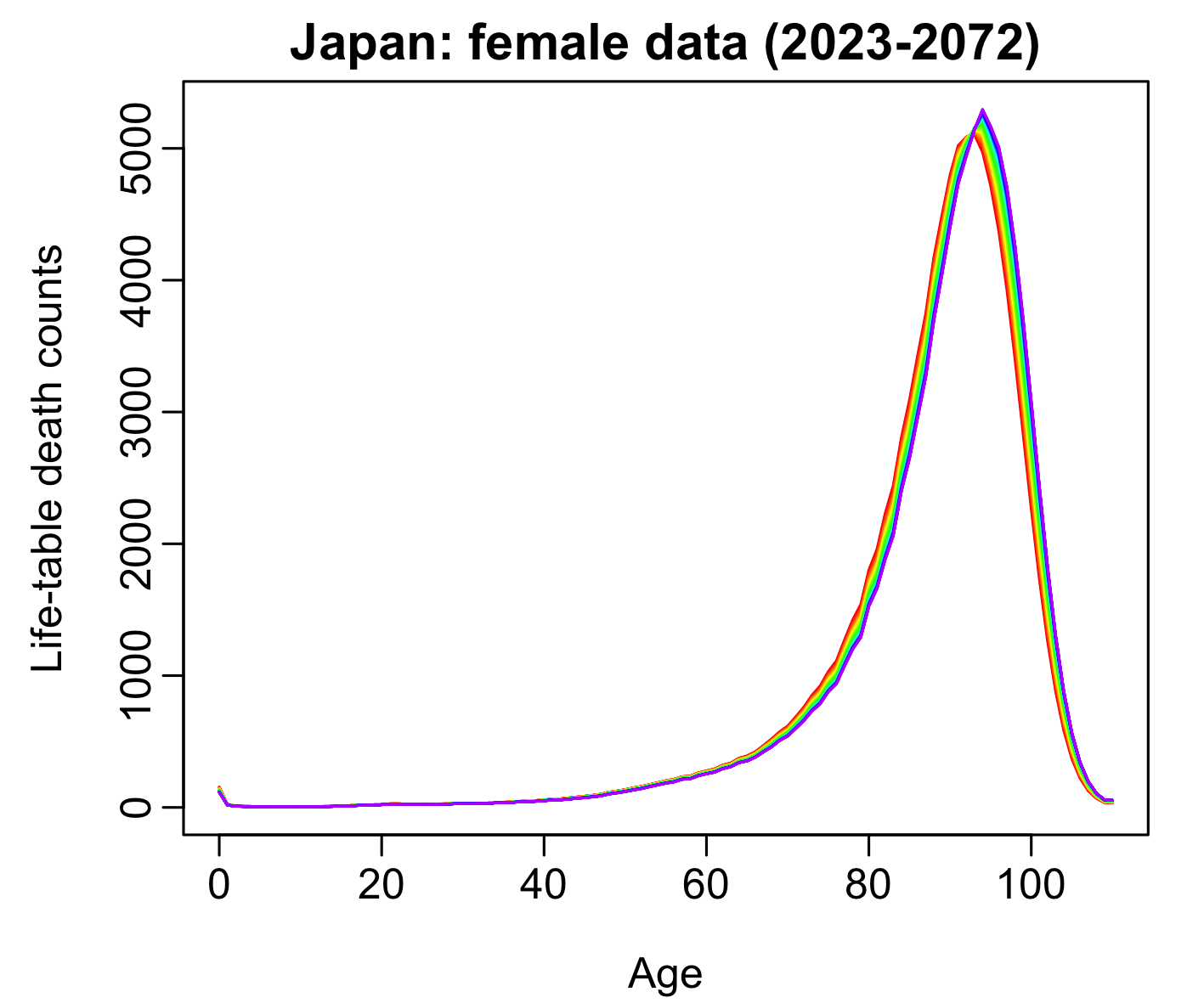}
\quad
\includegraphics[width=8.6cm]{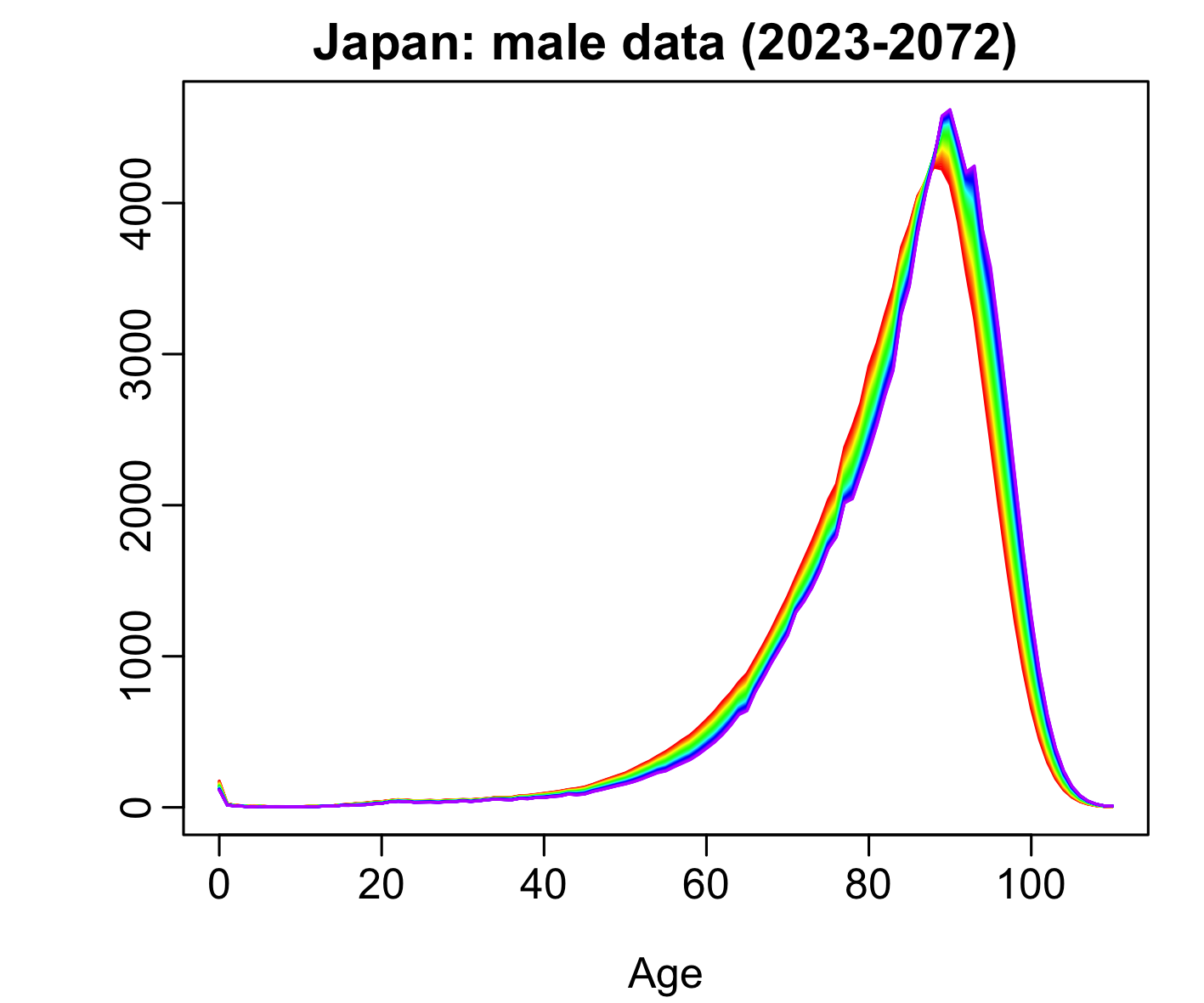}
\caption{\small{Age distribution of death count forecasts from 2023 to 2072 for Japanese females and males.}\label{fig:7}}
\end{figure}

Using the CDF transformation, we apply the multilevel functional time series method to obtain forecasts of age-specific life-table death counts and then determine the corresponding age-specific survival probabilities. In Figure~\ref{fig:7}, we display the age-specific life-table death count forecasts from 2023 to 2072 for Japanese females and males.

With the forecasts of life-table death counts, we input them into the calculation of single-premium term immediate annuities, and we adopt a cohort approach to calculating survival probabilities. The $\tau$ year survival probability of a person aged $x$ currently at $t=0$ is determined by
\begin{equation*}
_{\tau}p_x = \prod^{\tau}_{j=1}p_{x+j-1} = \prod^{\tau}_{j=1}(1-q_{x+j-1}) = \prod^{\tau}_{j=1}(1-\frac{d_{x+j-1}}{l_{x+j-1}}),
\end{equation*}
where $d_{x+j-1}$ denotes the number of death counts between ages $x+j-1$ and $x+j$ and $l_{x+j-1}$ denotes the number of lives alive at age $x+j-1$. 

The price of an annuity with a maturity term of a $T$ year is a random variable since it depends on the value of zero-coupon bond price and mortality forecasts. The annuity price written for an $x$-year-old with the benefit of one Japanese Yen per year is given by
\begin{equation*}
a_x^{T} = \sum^T_{\tau=1}B(0,\tau)_{\tau}p_{x},
\end{equation*}
where $B(0,\tau)$ is the $\tau$-year bond price and $_{\tau}p_x$ denotes the survival probability.

\begin{table}[!htb]
\centering
\tabcolsep 0.065in
\caption{\small{Estimates of annuity prices with different ages and maturities ($T$) for female and male populations residing in Japan. These estimates are based on forecast mortality from 2023 to 2072 and a constant interest rate $\eta=0.25\%$. We consider only contracts with maturity so that age + maturity $\leq 110$.}\label{tab:3}}
\begin{tabular}{@{}rrrrrrrrrrrrr@{}}
\toprule
	& \multicolumn{6}{c}{Female} 	& \multicolumn{6}{c}{Male} \\	
  Age & $T=5$ & 10 & 15 & 20 & 25 & 30 & 5 & 10 & 15 & 20 & 25 & 30 \\ 
\midrule
  60   & 4.916 & 9.671 & 14.215 & 18.467 & 22.274 & 25.364 & 4.859 & 9.430 & 13.606 & 17.258 & 20.224 & 22.285 \\ 
  65   & 4.895 & 9.574 & 13.951 & 17.871 & 21.052 & 23.133 & 4.804 & 9.193 & 13.032 & 16.150 & 18.316 & 19.418 \\ 
  70   & 4.855 & 9.397 & 13.464 & 16.766 & 18.924 & 19.851 & 4.712 & 8.832 & 12.178 & 14.504 & 15.687 & 16.050 \\ 
  75   & 4.784 & 9.067 & 12.544 & 14.818 & 15.794 & 15.996 & 4.578 & 8.295 & 10.878 & 12.192 & 12.596 & 12.652 \\ 
  80   & 4.635 & 8.398 & 10.859 & 11.915 & 12.133 & 12.150 & 4.352 & 7.376 & 8.914   & 9.387   & 9.453   & 9.457 \\ 
  85   & 4.324 & 7.152 & 8.366   & 8.617   & 8.636   &             & 3.901 & 5.885 & 6.494   & 6.580   & 6.585   &  \\ 
  90   & 3.720 & 5.316 & 5.647   & 5.672   &             &             & 3.187 & 4.165 & 4.303   & 4.310   &             &  \\ 
  95   & 2.829 & 3.415 & 3.461   &             &             &             & 2.335 & 2.664 & 2.682   &             &             &  \\ 
  100 & 1.851 & 1.993 &             &             &             &             & 1.551 & 1.636 &  	     &             &             &  \\ 
  105 & 1.176 &           &             &             &             &             & 1.043 &           &             &             &             &  \\ 
\bottomrule
\end{tabular}
\end{table}
In Table~\ref{tab:3}, to provide an example of the annuity calculations, we compute the point estimate of the annuity prices for various ages and maturities for female and male populations in Japan. We assume a constant current interest rate at $\eta=0.25\%$ (\url{https://tradingeconomics.com/japan/interest-rate}) and hence zero-coupon bond is given as $B(0,\tau) = \exp^{-\eta\tau}$. In Appendix~\ref{sec:appendix_D}, we present the best estimate of annuity prices under a constant interest rate of $3\%$. Although the difference in annuity price might appear small, any mispricing can result in a significant risk when considering a large contract value and annuity portfolio.

\section{Conclusion}\label{sec:6}

We introduce a CDF transformation to model and forecast the age distribution of deaths. Using the Japanese age- and sex-specific life-table death counts, we evaluate and compare the point and interval forecast accuracies between the CDF and clr transformations. Based on the error measures, the CDF transformation coupled with the MLFTS method is generally recommended for forecasting age distribution of deaths and the corresponding age-specific life expectancy. The superiority of this method is also driven by the ability to model female and male age distribution of deaths jointly. 

We apply the CDF transformation to forecast life-table death counts from 2023 to 2072. We then calculate the cumulative survival probability and obtain the best estimates of temporary annuity prices. As expected, we find that the cumulative survival probability has a pronounced impact on annuity prices. For reproducibility, the computer \Rlogo \ code is available at \url{https://github.com/hanshang/CDF_transformation}. 

There are a few ways in which this paper could be extended, and we briefly discussed four. 
\begin{enumerate}
\item[1)] A robust MLFTS method may be considered in the presence of outlying years. 
\item[2)] A model-averaging approach can improve accuracy for forecasting the age distribution of death counts.
\item[3)] The methodology can be applied to calculate other annuity prices, such as the whole-life immediate or deferred annuity. 
\item[4)] We may also consider a \textit{cohort} life table that depicts the life history of a specific group of individuals \citep[see, e.g.,][]{HH15, BKC20}, but is dependent on projected mortality rates for those cohorts born more recently. 
\end{enumerate}

\section*{Acknowledgements}

The first author acknowledges the financial support from an Australian Research Council Future Fellowship, FT240100338. 

\newpage
\appendix

\section{Additional results of point forecast accuracy comparison}\label{sec:appendix_A}

In Table~\ref{tab:Appendix_1}, we compare the point forecast errors between the CDF and clr transformations when the number of components is selected by the EVR criterion. As measured by the KLD and JSD, the CDF transformation coupled with the MLFTS method performs the best with the smallest averaged errors.
\begin{center}
\tabcolsep 0.15in
\renewcommand*{\arraystretch}{0.99}
\begin{longtable}{@{}llrrrrrrrr@{}}
\caption{\small{Comparison of the point forecast errors $(\times 100)$ between the CDF and clr transformations when the number of components is selected by the EVR criterion.}} \label{tab:Appendix_1} \\
\toprule
	& & \multicolumn{2}{c}{UFTS} & \multicolumn{2}{c}{MFTS} & \multicolumn{2}{c}{MLFTS} & \multicolumn{2}{c}{clr}  \\
Sex & $h$ & KLD & JSD & KLD & JSD & KLD & JSD & KLD & JSD \\ 
	\midrule
\endfirsthead
\toprule
	& & \multicolumn{2}{c}{UFTS} & \multicolumn{2}{c}{MFTS} & \multicolumn{2}{c}{MLFTS} & \multicolumn{2}{c}{clr} \\
Sex &  $h$ & KLD & JSD & KLD & JSD & KLD & JSD & KLD & JSD \\ 
		\midrule
\endhead	
\hline \multicolumn{10}{r}{{Continued on next page}} \\
\endfoot
\endlastfoot
F & 1 & 0.1011 & 0.0255 & 0.5286 & 0.1411 & \textBF{0.0863} & \textBF{0.0219} & 0.1524 & 0.0385 \\ 
   & 2 & 0.1649 & 0.0419 & 0.6040 & 0.1621 & \textBF{0.1277} & \textBF{0.0325} & 0.2389 & 0.0605 \\ 
   & 3 & 0.2410 & 0.0611 & 0.6589 & 0.1771 & \textBF{0.1443} & \textBF{0.0365} & 0.3365 & 0.0850 \\ 
   & 4 & 0.3594 & 0.0915 & 0.7364 & 0.1988 & \textBF{0.1906} & \textBF{0.0482} & 0.4785 & 0.1210 \\ 
   & 5 & 0.4908 & 0.1252 & 0.7784 & 0.2110 & \textBF{0.2535} & \textBF{0.0640} & 0.6321 & 0.1598 \\ 
   & 6 & 0.6162 & 0.1573 & 0.7728 & 0.2099 & \textBF{0.2889} & \textBF{0.0729} & 0.7832 & 0.1980 \\ 
   & 7 & 0.7847 & 0.1993 & 0.7584 & 0.2061 & \textBF{0.3512} & \textBF{0.0880} & 1.0116 & 0.2535 \\ 
   & 8 & 1.0604 & 0.2672 & 0.7425 & 0.2014 & \textBF{0.3394} & \textBF{0.0855} & 1.2974 & 0.3225 \\ 
   & 9 & 1.3333 & 0.3363 & 0.7935 & 0.2155 & \textBF{0.3098} & \textBF{0.0791} & 1.5357 & 0.3815 \\ 
   & 10 & 1.3743 & 0.3539 & 0.7971 & 0.2163 & \textBF{0.3083} & \textBF{0.0790} & 1.5183 & 0.3823 \\ 
   & 11 & 1.2208 & 0.3236 & 0.8236 & 0.2226 & \textBF{0.3827} & \textBF{0.0990} & 1.2251 & 0.3155 \\ 
   & 12 & 1.2978 & 0.3568 & 1.1603 & 0.3199 & \textBF{0.4369} & \textBF{0.1126} & 1.0567 & 0.2824 \\ 
   & 13 & 1.8033 & 0.5035 & 1.2738 & 0.3536 & \textBF{0.5649} & \textBF{0.1475} & 1.4206 & 0.3817 \\ 
   & 14 & 2.1521 & 0.6062 & 1.3081 & 0.3635 & \textBF{0.5291} & \textBF{0.1368} & 1.7751 & 0.4758 \\ 
   & 15 & 3.2044 & 0.9241 & 1.8190 & 0.5143 & \textBF{0.8501} & \textBF{0.2245} & 2.8910 & 0.7893 \\ 
   & 16 & 4.1412 & 1.2129 & 2.0189 & 0.5716 & \textBF{1.9506} & \textBF{0.5419} & 4.1828 & 1.1635 \\ 
  \cmidrule{2-10}
   & Mean & 1.2716 & 0.3491 & 0.9734 & 0.2678 & \textBF{0.4446} & \textBF{0.1169} & 1.2835 & 0.3382 \\ 
  \cmidrule{2-10}
  M & 1 & 0.0983 & 0.0248 & 0.1563 & 0.0401 & \textBF{0.0973} & \textBF{0.0245} & 0.1919 & 0.0480 \\ 
      & 2 & 0.1328 & 0.0334 & 0.1923 & 0.0494 & \textBF{0.1285} & \textBF{0.0323} & 0.2710 & 0.0676 \\ 
      & 3 & 0.1718 & 0.0430 & 0.2268 & 0.0582 & \textBF{0.1588} & \textBF{0.0398} & 0.3579 & 0.0888 \\ 
      & 4 & 0.2196 & 0.0550 & 0.2728 & 0.0700 & \textBF{0.2026} & \textBF{0.0509} & 0.4641 & 0.1149 \\ 
      & 5 & 0.2774 & 0.0694 & 0.3164 & 0.0812 & \textBF{0.2572} & \textBF{0.0646} & 0.5811 & 0.1434 \\ 
      & 6 & 0.3530 & 0.0881 & 0.3584 & 0.0919 & \textBF{0.3213} & \textBF{0.0806} & 0.7151 & 0.1759 \\ 
      & 7 & 0.4318 & 0.1074 & 0.4015 & 0.1028 & \textBF{0.3788} & \textBF{0.0945} & 0.8803 & 0.2148 \\ 
      & 8 & 0.5290 & 0.1315 & \textBF{0.4478} & 0.1144 & 0.4497 & \textBF{0.1122} & 1.0840 & 0.2614 \\ 
      & 9 & 0.6148 & 0.1535 & 0.4962 & 0.1266 & \textBF{0.4929} & \textBF{0.1230} & 1.2672 & 0.3035 \\ 
      & 10 & 0.6564 & 0.1652 & 0.5062 & 0.1294 & \textBF{0.4918} & \textBF{0.1234} & 1.2893 & 0.3123 \\ 
      & 11 & 0.6125 & 0.1557 & 0.4831 & 0.1237 & \textBF{0.3435} & \textBF{0.0891} & 0.9209 & 0.2268 \\ 
      & 12 & 0.5962 & 0.1519 & 0.5057 & 0.1301 & \textBF{0.3467} & \textBF{0.0902} & 0.3806 & 0.0957 \\ 
      & 13 & 0.6304 & 0.1613 & 0.5326 & 0.1376 & \textBF{0.3710} & 0.0975 & 0.3724 & \textBF{0.0937} \\ 
      & 14 & 0.6669 & 0.1711 & 0.5431 & 0.1402 & 0.3699 & 0.0979 & \textBF{0.3355} & \textBF{0.0848} \\ 
      & 15 & 0.6473 & 0.1673 & 0.5767 & 0.1509 & 0.5385 & 0.1457 & \textBF{0.2462} & \textBF{0.0630} \\ 
      & 16 & 0.5901 & 0.1537 & 0.5596 & 0.1473 & 0.9476 & 0.2628 & \textBF{0.2468} & \textBF{0.0653} \\ 
  \cmidrule{2-10}
     & Mean & 0.4518 & 0.1145 & 0.4110 & 0.1059 & \textBF{0.3685} & \textBF{0.0956} & 0.6003 & 0.1475 \\ 
\bottomrule
\end{longtable}
\end{center}   

\vspace{-.5in}

In Table~\ref{tab:e0_2}, we compare the point forecast errors (as measured by RMSFE and MAFE) of the age-specific life expectancy between the CDF and clr transformation under the EVR criterion. The CDF transformation, coupled with the MLFTS method, gives the smallest forecast errors.
\begin{center}
\tabcolsep 0.125in
\renewcommand*{\arraystretch}{0.99}
\begin{longtable}{@{}llrrrrrrrr@{}}
\caption{\small{Comparison of the point forecast errors of the life expectancy between the CDF and clr transformations, when the number of components is selected by the EVR criterion.}} \label{tab:e0_2} \\
\toprule
	& & \multicolumn{2}{c}{UFTS} & \multicolumn{2}{c}{MFTS} & \multicolumn{2}{c}{MLFTS} & \multicolumn{2}{c}{clr}  \\
Sex & $h$ & RMSFE & MAFE & RMSFE & MAFE & RMSFE & MAFE & RMSFE & MAFE \\ 
	\midrule
\endfirsthead
\toprule
	& & \multicolumn{2}{c}{UFTS} & \multicolumn{2}{c}{MFTS} & \multicolumn{2}{c}{MLFTS} & \multicolumn{2}{c}{clr} \\
Sex &  $h$ & RMSFE & MAFE & RMSFE & MAFE & RMSFE & MAFE & RMSFE & MAFE \\ 
		\midrule
\endhead	
\hline \multicolumn{10}{r}{{Continued on next page}} \\
\endfoot
\endlastfoot
F & 1 & 0.2682 & 0.1967 & 0.5308 & 0.4200 & \textBF{0.2231} & \textBF{0.1736} & 0.3551 & 0.2779 \\ 
  & 2 & 0.3410 & 0.2346 & 0.5630 & 0.4416 & \textBF{0.2826} & \textBF{0.2191} & 0.4607 & 0.3637 \\ 
  & 3 & 0.4157 & 0.2898 & 0.5761 & 0.4703 & \textBF{0.2842} & \textBF{0.2202} & 0.5608 & 0.4324 \\ 
  & 4 & 0.5179 & 0.3669 & 0.6045 & 0.4920 & \textBF{0.3333} & \textBF{0.2586} & 0.6870 & 0.5321 \\ 
  & 5 & 0.6072 & 0.4358 & 0.6064 & 0.4968 & \textBF{0.3987} & \textBF{0.3272} & 0.8003 & 0.6251 \\ 
  & 6 & 0.6727 & 0.4818 & 0.5728 & 0.4628 & \textBF{0.4095} & \textBF{0.3102} & 0.8896 & 0.6969 \\ 
  & 7 & 0.7669 & 0.5696 & 0.5407 & 0.4323 & \textBF{0.4863} & \textBF{0.3920} & 1.0176 & 0.7927 \\ 
  & 8 & 0.9082 & 0.6603 & 0.5043 & 0.4149 & \textBF{0.4628} & \textBF{0.3614} & 1.1597 & 0.9066 \\ 
  & 9 & 1.0160 & 0.7600 & 0.5088 & 0.4446 & \textBF{0.4111} & \textBF{0.3412} & 1.2673 & 1.0041 \\ 
  & 10 & 0.9950 & 0.7612 & 0.4736 & 0.4184 & \textBF{0.3350} & \textBF{0.2842} & 1.2661 & 1.0068 \\ 
  & 11 & 0.8346 & 0.6767 & 0.4471 & 0.4188 & \textBF{0.3282} & \textBF{0.2755} & 1.1243 & 0.9277 \\ 
  & 12 & 0.7088 & 0.6739 & 0.6328 & 0.5486 & \textBF{0.3454} & \textBF{0.2885} & 0.9909 & 0.9087 \\ 
  & 13 & 0.8919 & 0.8341 & 0.6776 & 0.5731 & \textBF{0.3651} & \textBF{0.3169} & 1.1649 & 1.0740 \\ 
  & 14 & 1.0003 & 0.9352 & 0.6847 & 0.5788 & \textBF{0.2570} & \textBF{0.1972} & 1.3319 & 1.2290 \\ 
  & 15 & 1.2775 & 1.1902 & 0.8604 & 0.7441 & \textBF{0.3391} & \textBF{0.3009} & 1.6866 & 1.5502 \\ 
  & 16 & 1.4662 & 1.4099 & 0.8776 & 0.8514 & \textBF{0.8056} & \textBF{0.7786} & 1.9981 & 1.8592 \\ 
\cmidrule{2-10}  
  & Mean & 0.7930 & 0.6548 & 0.6038 & 0.5130 & \textBF{0.3792} & \textBF{0.3153} & 1.0476 & 0.8867 \\
\midrule
M &  1 & 0.2597 & 0.2116 & 0.2852 & 0.2209 & \textBF{0.2101} & \textBF{0.1201} & 0.4447 & 0.3361 \\ 
  & 2 & 0.3028 & 0.2512 & 0.3380 & 0.2605 & \textBF{0.2373} & \textBF{0.1506} & 0.5448 & 0.4084 \\ 
  & 3 & 0.3516 & 0.2899 & 0.3856 & 0.2931 & \textBF{0.2597} & \textBF{0.1721} & 0.6407 & 0.4573 \\ 
  & 4 & 0.4054 & 0.3298 & 0.4415 & 0.3400 & \textBF{0.3000} & \textBF{0.2195} & 0.7420 & 0.5138 \\ 
  & 5 & 0.4701 & 0.3777 & 0.4974 & 0.3922 & \textBF{0.3546} & \textBF{0.2720} & 0.8428 & 0.5894 \\ 
  & 6 & 0.5429 & 0.4422 & 0.5555 & 0.4509 & \textBF{0.4083} & \textBF{0.3149} & 0.9455 & 0.6657 \\ 
  & 7 & 0.6127 & 0.4910 & 0.6137 & 0.4999 & \textBF{0.4539} & \textBF{0.3633} & 1.0582 & 0.7498 \\ 
  & 8 & 0.6956 & 0.5553 & 0.6664 & 0.5372 & \textBF{0.5178} & \textBF{0.4233} & 1.1791 & 0.8306 \\ 
  & 9 & 0.7952 & 0.6374 & 0.7103 & 0.5757 & \textBF{0.5914} & \textBF{0.4869} & 1.2803 & 0.8994 \\ 
  & 10 & 0.8703 & 0.7289 & 0.7219 & 0.5946 & \textBF{0.6313} & \textBF{0.5057} & 1.3014 & 0.9483 \\ 
  & 11 & 0.8723 & 0.7369 & 0.6997 & 0.5893 & \textBF{0.4865} & \textBF{0.3936} & 1.0912 & 0.7971 \\ 
  & 12 & 0.8512 & 0.7069 & 0.6846 & 0.5588 & \textBF{0.4453} & \textBF{0.3665} & 0.6330 & 0.5167 \\ 
  & 13 & 0.8529 & 0.6996 & 0.7035 & 0.5708 & \textBF{0.4059} & \textBF{0.3321} & 0.6240 & 0.5190 \\ 
  & 14 & 0.8601 & 0.6916 & 0.7061 & 0.5677 & \textBF{0.3328} & \textBF{0.2808} & 0.5718 & 0.4672 \\ 
  & 15 & 0.7898 & 0.6364 & 0.6313 & 0.5049 & \textBF{0.3066} & \textBF{0.2528} & 0.4091 & 0.3462 \\ 
  & 16 & 0.6301 & 0.5333 & \textBF{0.4684} & \textBF{0.4113} & 0.4724 & 0.4278 & 0.1757 & 0.1455 \\ 
\cmidrule{2-10}
  & Mean & 0.6352 & 0.5200 & 0.5693 & 0.4605 & \textBF{0.4009} & \textBF{0.3176} & 0.7803 & 0.5744 \\ 
\bottomrule
\end{longtable}
\end{center}   
   
\section{Additional results of interval forecast accuracy comparison}\label{sec:appendix_B}

In Table~\ref{tab:Appendix_2}, we present the comparison of the interval forecast accuracy between the CDF and clr transformation at the 80\% nominal coverage probability when the number of components is selected by the EVR criterion. For the female data, the clr transformation produces the smallest averaged interval score, while the CDF transformation coupled with the MLFTS method gives the smallest averaged CPD. For the male data, the clr transformation gives the smallest interval forecast errors. The mixed results motivate us to consider a model-averaging approach by assigning weights to forecasts obtained from the two transformations.
\begin{center}
\renewcommand*{\arraystretch}{0.96}
\tabcolsep 0.15in
\begin{longtable}{@{}llrrrrrrrr@{}}
\caption{\small{Comparison of the interval forecast errors between the CDF and clr transformations at the 80\% nominal coverage probability when the number of components is selected by the EVR criterion.}} \label{tab:Appendix_2} \\
\toprule
	& & \multicolumn{2}{c}{UFTS} & \multicolumn{2}{c}{MFTS} & \multicolumn{2}{c}{MLFTS} & \multicolumn{2}{c}{clr}  \\
Sex & $h$ & score & CPD & score & CPD & score & CPD & score & CPD  \\ 
	\midrule
\endfirsthead
\toprule
	& & \multicolumn{2}{c}{UFTS} & \multicolumn{2}{c}{MFTS} & \multicolumn{2}{c}{MLFTS} & \multicolumn{2}{c}{clr} \\
Sex & $h$ & score & CPD & score & CPD & score & CPD & score & CPD \\ 
	\midrule
\endhead
\hline \multicolumn{10}{r}{{Continued on next page}} \\
\endfoot
\endlastfoot
F & 1 & 0.0037 & 0.1842 & 0.0019 & 0.1482 & \textBF{0.0017} & 0.0654 & 0.0017 & \textBF{0.0153} \\ 
   & 2 & 0.0039 & 0.1820 & 0.0021 & 0.1279 & 0.0021 & 0.0517 & \textBF{0.0020} & \textBF{0.0006} \\ 
   & 3 & 0.0041 & 0.1910 & \textBF{0.0023} & 0.1337 & 0.0024 & 0.0629 & 0.0024 & \textBF{0.0385} \\ 
   & 4 & 0.0043 & 0.1931 & \textBF{0.0027} & 0.1272 & 0.0029 & 0.0621 & 0.0027 & \textBF{0.0219} \\ 
   & 5 & 0.0046 & 0.1925 & \textBF{0.0030} & 0.1092 & 0.0033 & \textBF{0.0363} & 0.0031 & 0.0544 \\ 
   & 6 & 0.0049 & 0.1943 & \textBF{0.0034} & 0.1247 & 0.0039 & \textBF{0.0943} & 0.0034 & 0.1181 \\ 
   & 7 & 0.0051 & 0.1964 & 0.0038 & 0.1261 & 0.0042 & \textBF{0.1000} & \textBF{0.0036} & 0.1162 \\ 
   & 8 & 0.0054 & 0.1970 & 0.0043 & 0.1239 & 0.0048 & \textBF{0.1089} & \textBF{0.0042} & 0.1299 \\ 
   & 9 & 0.0055 & 0.1944 & 0.0047 & 0.1144 & 0.0051 & \textBF{0.1020} & \textBF{0.0044} & 0.1572 \\ 
   & 10 & 0.0059 & 0.1858 & 0.0053 & 0.1254 & 0.0055 & \textBF{0.1215} & \textBF{0.0049} & 0.1562 \\ 
   & 11 & 0.0061 & 0.1895 & 0.0059 & 0.1264 & 0.0060 & \textBF{0.1249} & \textBF{0.0052} & 0.1655 \\ 
   & 12 & 0.0067 & 0.1802 & 0.0065 & 0.1477 & 0.0069 & \textBF{0.1207} & \textBF{0.0059} & 0.1586 \\ 
   & 13 & 0.0076 & 0.1595 & 0.0073 & 0.1392 & 0.0078 & \textBF{0.1099} & \textBF{0.0070} & 0.1662 \\ 
   & 14 & 0.0082 & 0.1610 & 0.0081 & 0.1369 & 0.0082 & \textBF{0.1069} & \textBF{0.0074} & 0.1610 \\ 
   & 15 & 0.0092 & 0.1550 & 0.0097 & 0.1369 & 0.0090 & \textBF{0.0784} & \textBF{0.0083} & 0.1640 \\ 
   & 16 & 0.0103 & 0.1550 & \textBF{0.0097} & 0.1459 & 0.0102 & \textBF{0.0018} & 0.0097 & 0.1550 \\ 
\cmidrule{2-10}
   & Mean & 0.0060 & 0.1819 & 0.0050 & 0.1309 & 0.0052 & \textBF{0.0842} & \textBF{0.0047} & 0.1112 \\ 
\cmidrule{2-10}
M & 1 & 0.0041 & 0.1803 & 0.0019 & 0.1583 & \textBF{0.0015} & 0.0722 & 0.0017 & \textBF{0.0367} \\ 
  & 2 & 0.0042 & 0.1784 & 0.0020 & 0.1568 & \textBF{0.0018} & 0.0703 & 0.0019 & \textBF{0.0505} \\ 
  & 3 & 0.0043 & 0.1910 & 0.0021 & 0.1537 & \textBF{0.0021} & 0.0835 & 0.0023 & \textBF{0.0346} \\ 
  & 4 & 0.0045 & 0.1986 & \textBF{0.0022} & 0.1626 & 0.0024 & 0.0829 & 0.0026 & \textBF{0.0420} \\ 
  & 5 & 0.0047 & 0.2000 & \textBF{0.0024} & 0.1512 & 0.0027 & 0.0679 & 0.0029 & \textBF{0.0551} \\ 
  & 6 & 0.0049 & 0.2000 & \textBF{0.0026} & 0.1271 & 0.0031 & 0.0935 & 0.0033 & \textBF{0.0452} \\ 
  & 7 & 0.0051 & 0.1964 & \textBF{0.0029} & 0.1297 & 0.0033 & 0.0766 & 0.0036 & \textBF{0.0252} \\ 
  & 8 & 0.0053 & 0.2000 & \textBF{0.0032} & 0.1179 & 0.0038 & 0.0729 & 0.0040 & \textBF{0.0238} \\ 
  & 9 & 0.0054 & 0.2000 & \textBF{0.0034} & 0.0953 & 0.0038 & 0.0716 & 0.0042 & \textBF{0.0232} \\ 
  & 10 & 0.0055 & 0.2000 & 0.0038 & 0.0867 & 0.0040 & 0.0945 & \textBF{0.0037} & \textBF{0.0520} \\ 
  & 11 & 0.0055 & 0.2000 & 0.0041 & \textBF{0.0664} & 0.0041 & 0.1204 & \textBF{0.0034} & 0.1024 \\ 
  & 12 & 0.0057 & 0.2000 & 0.0043 & \textBF{0.0541} & 0.0045 & 0.1441 & \textBF{0.0040} & 0.1820 \\ 
  & 13 & 0.0061 & 0.2000 & 0.0047 & \textBF{0.0514} & 0.0049 & 0.1482 & \textBF{0.0043} & 0.1865 \\ 
  & 14 & 0.0061 & 0.2000 & 0.0050 & \textBF{0.0739} & 0.0053 & 0.1489 & \textBF{0.0042} & 0.1880 \\ 
  & 15 & 0.0065 & 0.2000 & 0.0057 & \textBF{0.0919} & 0.0057 & 0.1414 & \textBF{0.0047} & 0.1910 \\ 
  & 16 & 0.0068 & 0.2000 & 0.0058 & \textBF{0.1099} & 0.0064 & 0.1550 & \textBF{0.0053} & 0.1820 \\ 
\cmidrule{2-10}
  & Mean & 0.0053 & 0.1965 & 0.0035 & 0.1117 & 0.0037 & 0.1027 & \textBF{0.0035} & \textBF{0.0888} \\ 
\bottomrule
\end{longtable}
\end{center}   

\vspace{-.7in}

\section{Comparison of the interval forecast errors between the CDF and clr transformations at the \mbox{95\% nominal coverage probability}}\label{sec:appendix_C}

In Table~\ref{tab:Appendix_3}, we present the comparison of the interval forecast accuracy between the CDF and clr transformation at the 95\% nominal coverage probability. We consider two ways of selecting the number of components: EVR criterion or $K=6$. When the number of components is selected by the EVR criterion, the CDF transformation with the MFTS method gives the smallest average scores, while the CPD transformation with the MLFTS method produces the smallest averaged CPD for the female and male series. When the number of components is fixed at $K=6$, the CDF transformation with the UFTS method produces the smallest averaged interval scores and CPD for the female series. For the male series, the CDF transformation with the MLFTS method gives the smallest averaged interval scores and CPD.

\begin{center}
\tabcolsep 0.1in
\begin{longtable}{@{}lllrrrrrrrr@{}}
\caption{\small{Comparison of the interval forecast errors between the CDF and clr transformation at the 95\% nominal coverage probability. The number of components is fixed at $K=6$ or selected by the EVR criterion.}} \label{tab:Appendix_3} \\
\toprule
	& & & \multicolumn{2}{c}{UFTS} & \multicolumn{2}{c}{MFTS} & \multicolumn{2}{c}{MLFTS} & \multicolumn{2}{c}{clr} \\
Criterion & Sex & $h$ & score & CPD & score & CPD & score & CPD & score & CPD \\ 
	\midrule
\endfirsthead
\toprule
	& & & \multicolumn{2}{c}{UFTS} & \multicolumn{2}{c}{MFTS} & \multicolumn{2}{c}{MLFTS} & \multicolumn{2}{c}{clr} \\
Criterion & Sex & $h$ & score & CPD & score & CPD & score & CPD & score & CPD\\ 
	\midrule
\endhead
\hline \multicolumn{11}{r}{{Continued on next page}} \\
\endfoot
\endlastfoot
EVR & F 	& 1    & 0.0064 & 0.0477 & 0.0028 & 0.0382 & \textBF{0.0025} & \textBF{0.0021} & 0.0025 & 0.0226 \\ 
  		& & 2 & 0.0065 & 0.0494 & 0.0030 & 0.0368 & 0.0029 & \textBF{0.0047} & \textBF{0.0028} & 0.0161 \\ 
		& & 3 & 0.0069 & 0.0500 & \textBF{0.0034} & 0.0378 & 0.0035 & \textBF{0.0034} & 0.0034 & 0.0144 \\ 
  		& & 4 & 0.0071 & 0.0493 & 0.0039 & 0.0368 & 0.0041 & 0.0103 & \textBF{0.0039} & \textBF{0.0082} \\ 
  		& & 5 & 0.0075 & 0.0500 & 0.0044 & 0.0335 & 0.0046 & 0.0146 & \textBF{0.0044} & \textBF{0.0072} \\ 
  		& & 6 & 0.0082 & 0.0500 & \textBF{0.0051} & 0.0353 & 0.0058 & \textBF{0.0032} & 0.0056 & 0.0312 \\ 
  		& & 7 & 0.0081 & 0.0500 & \textBF{0.0056} & 0.0320 & 0.0064 & \textBF{0.0041} & 0.0057 & 0.0338 \\ 
  		& & 8 & 0.0086 & 0.0500 & \textBF{0.0061} & 0.0330 & 0.0071 & \textBF{0.0100} & 0.0063 & 0.0400 \\ 
  		& & 9 & 0.0091 & 0.0500 & \textBF{0.0066} & 0.0354 & 0.0082 & \textBF{0.0173} & 0.0074 & 0.0399 \\ 
  		& & 10 & 0.0095 & 0.0500 & \textBF{0.0072} & 0.0307 & 0.0090 & \textBF{0.0243} & 0.0083 & 0.0474 \\ 
  		& & 11 & 0.0103 & 0.0500 & \textBF{0.0080} & 0.0380 & 0.0105 & \textBF{0.0260} & 0.0097 & 0.0485 \\ 
  		& & 12 & 0.0103 & 0.0500 & \textBF{0.0088} & 0.0356 & 0.0116 & \textBF{0.0212} & 0.0095 & 0.0464 \\ 
  		& & 13 & 0.0116 & 0.0500 & \textBF{0.0101} & 0.0252 & 0.0130 & \textBF{0.0095} & 0.0113 & 0.0410 \\ 
  		& & 14 & 0.0122 & 0.0500 & \textBF{0.0112} & 0.0170 & 0.0140 & \textBF{0.0140} & 0.0119 & 0.0350 \\ 
  		& & 15 & 0.0132 & 0.0500 & \textBF{0.0127} & 0.0185 & 0.0147 & \textBF{0.0050} & 0.0133 & 0.0275 \\ 
  		& & 16 & \textBF{0.0141} & 0.0500 & 0.0148 & 0.0050 & 0.0163 & \textBF{0.0041} & 0.0151 & 0.0230 \\ 
\cmidrule{2-11}
  		& & Mean & 0.0094 & 0.0498 & \textBF{0.0071} & 0.0305 & 0.0084 & \textBF{0.0108} & 0.0076 & 0.0301 \\ 
\cmidrule{2-11}
& M  & 1    & 0.0062 & 0.0444 & 0.0029 & 0.0348 & \textBF{0.0022} & 0.0168 & 0.0022 & \textBF{0.0162} \\ 
	& & 2 & 0.0063 & 0.0470 & 0.0030 & 0.0320 & \textBF{0.0025} & 0.0230 & 0.0025 & \textBF{0.0200} \\ 
	& & 3 & 0.0065 & 0.0494 & 0.0030 & 0.0455 & \textBF{0.0028} & 0.0268 & 0.0029 & \textBF{0.0172} \\ 
	& & 4 & 0.0067 & 0.0500 & 0.0033 & 0.0465 & \textBF{0.0032} & 0.0278 & 0.0035 & \textBF{0.0061} \\ 
	& & 5 & 0.0070 & 0.0500 & 0.0035 & 0.0477 & \textBF{0.0035} & 0.0267 & 0.0037 & \textBF{0.0155} \\ 
	& & 6 & 0.0075 & 0.0500 & \textBF{0.0039} & 0.0492 & 0.0044 & 0.0361 & 0.0044 & \textBF{0.0344} \\ 
	& & 7 & 0.0075 & 0.0500 & \textBF{0.0041} & 0.0491 & 0.0047 & 0.0311 & 0.0046 & \textBF{0.0014} \\ 
	& & 8 & 0.0080 & 0.0500 & \textBF{0.0044} & 0.0500 & 0.0051 & 0.0290 & 0.0050 & \textBF{0.0090} \\ 
	& & 9 & 0.0083 & 0.0500 & \textBF{0.0047} & 0.0455 & 0.0058 & 0.0230 & 0.0058 & \textBF{0.0018} \\ 
	& & 10 & 0.0086 & 0.0500 & \textBF{0.0050} & 0.0423 & 0.0062 & \textBF{0.0217} & 0.0061 & 0.0358 \\ 
	& & 11 & 0.0090 & 0.0500 & \textBF{0.0055} & 0.0410 & 0.0070 & \textBF{0.0305} & 0.0069 & 0.0470 \\ 
	& & 12 & 0.0090 & 0.0500 & \textBF{0.0057} & 0.0392 & 0.0073 & \textBF{0.0320} & 0.0066 & 0.0482 \\ 
	& & 13 & 0.0096 & 0.0500 & \textBF{0.0061} & 0.0410 & 0.0077 & \textBF{0.0297} & 0.0075 & 0.0477 \\ 
	& & 14 & 0.0098 & 0.0500 & \textBF{0.0064} & 0.0440 & 0.0083 & \textBF{0.0230} & 0.0075 & 0.0440 \\ 
	& & 15 & 0.0103 & 0.0500 & \textBF{0.0069} & 0.0320 & 0.0086 & \textBF{0.0185} & 0.0081 & 0.0455 \\ 
	& & 16 & 0.0109 & 0.0500 & \textBF{0.0075} & 0.0230 & 0.0097 & \textBF{0.0050} & 0.0090 & 0.0410 \\ 
\cmidrule{2-11}
	& & Mean   & 0.0082 & 0.0494 & \textBF{0.0047} & 0.0414 & 0.0056 & \textBF{0.0250} & 0.0054 & 0.0269 \\   
\midrule
$K=6$ & F  & 1  & 0.0023 & \textBF{0.0151} & 0.0028 & 0.0370 & \textBF{0.0018} & 0.0429 & 0.0021 & 0.0271 \\ 
	 	  & & 2  & 0.0025 & \textBF{0.0068} & 0.0030 & 0.0344 & \textBF{0.0021} & 0.0419 & 0.0023 & 0.0323 \\ 
		  & & 3  & 0.0030 & \textBF{0.0037} & 0.0034 & 0.0384 & \textBF{0.0026} & 0.0562 & 0.0029 & 0.0189 \\ 
   		  & & 4  & 0.0035 & \textBF{0.0027} & 0.0040 & 0.0361 & \textBF{0.0031} & 0.0567 & 0.0035 & 0.0179 \\ 
		  & & 5  & 0.0039 & 0.0168 & 0.0045 & 0.0305 & \textBF{0.0036} & 0.0784 & 0.0038 & \textBF{0.0108} \\ 
 		  & & 6  & 0.0046 & \textBF{0.0081} & 0.0050 & 0.0361 & \textBF{0.0043} & 0.0385 & 0.0046 & 0.0140 \\ 
		  & & 7  & 0.0047 & \textBF{0.0050} & 0.0055 & 0.0293 & \textBF{0.0043} & 0.0356 & 0.0046 & 0.0239 \\ 
  		  & & 8  & 0.0053 & \textBF{0.0061} & 0.0061 & 0.0350 & \textBF{0.0051} & 0.0211 & 0.0053 & 0.0350 \\ 
		  & & 9  & 0.0055 & \textBF{0.0016} & 0.0067 & 0.0376 & 0.0053 & 0.0311 & \textBF{0.0053} & 0.0387 \\ 
 		  & & 10 & 0.0059 & \textBF{0.0101} & 0.0072 & 0.0333 & 0.0060 & 0.0195 & \textBF{0.0058} & 0.0384 \\ 
		  & & 11 & 0.0064 & 0.0170 & 0.0081 & 0.0365 & 0.0066 & \textBF{0.0131} & \textBF{0.0063} & 0.0410 \\ 
		  & & 12 & 0.0074 & 0.0248 & 0.0090 & 0.0356 & \textBF{0.0072} & \textBF{0.0032} & 0.0073 & 0.0374 \\ 
		  & & 13 & 0.0087 & 0.0117 & 0.0102 & 0.0230 & \textBF{0.0085} & \textBF{0.0095} & 0.0086 & 0.0252 \\ 
		  & & 14 & 0.0096 & 0.0170 & 0.0111 & 0.0200 & \textBF{0.0092} & \textBF{0.0020} & 0.0098 & 0.0170 \\ 
		  & & 15 & \textBF{0.0110} & \textBF{0.0005} & 0.0127 & 0.0140 & 0.0110 & 0.0221 & 0.0112 & 0.0095 \\ 
		  & & 16 & \textBF{0.0127} & \textBF{0.0050} & 0.0135 & 0.0050 & 0.0165 & 0.1302 & 0.0138 & 0.0050 \\ 
\cmidrule{2-11}
		  & & Mean & \textBF{0.0061} & \textBF{0.0095} & 0.0070 & 0.0301 & 0.0061 & 0.0376 & 0.0061 & 0.0245 \\ 
\midrule
& M 	   	& 1  & 0.0024 & 0.0370 & 0.0029 & 0.0359 & \textBF{0.0016} & 0.0176 & 0.0019 & \textBF{0.0089} \\ 
    	    & & 2   & 0.0026 & 0.0350 & 0.0029 & 0.0332 & \textBF{0.0019} & 0.0233 & 0.0022 & \textBF{0.0164} \\ 
	    & & 3   & 0.0029 & 0.0436 & 0.0031 & 0.0423 & \textBF{0.0021} & \textBF{0.0079} & 0.0027 & 0.0140 \\ 
	    & & 4   & 0.0032 & 0.0375 & 0.0033 & 0.0500 & \textBF{0.0025} & \textBF{0.0103} & 0.0030 & 0.0202 \\ 
	    & & 5   & 0.0036 & 0.0312 & 0.0036 & 0.0500 & \textBF{0.0028} & 0.0168 & 0.0032 & \textBF{0.0080} \\ 
	    & & 6   & 0.0041 & 0.0320 & 0.0038 & 0.0459 & \textBF{0.0034} & \textBF{0.0049} & 0.0038 & 0.0181 \\ 
	    & & 7   & 0.0044 & 0.0248 & 0.0041 & 0.0482 & \textBF{0.0037} & 0.0167 & 0.0040 & \textBF{0.0113} \\ 
	    & & 8   & 0.0051 & 0.0230 & 0.0044 & 0.0470 & \textBF{0.0041} & \textBF{0.0081} & 0.0046 & 0.0150 \\ 
	    & & 9   & 0.0053 & 0.0140 & 0.0047 & 0.0444 & \textBF{0.0043} & 0.0041 & 0.0047 & \textBF{0.0027} \\ 
	    & & 10 & 0.0056 & 0.0165 & 0.0050 & 0.0397 & \textBF{0.0043} & 0.0191 & 0.0045 & \textBF{0.0024} \\ 
	    & & 11 & 0.0054 & \textBF{0.0080} & 0.0054 & 0.0425 & \textBF{0.0045} & 0.0140 & 0.0046 & 0.0200 \\ 
	    & & 12 & 0.0059 & \textBF{0.0023} & 0.0059 & 0.0374 & \textBF{0.0048} & 0.0140 & 0.0050 & 0.0338 \\ 
	    & & 13 & 0.0065 & \textBF{0.0018} & 0.0062 & 0.0410 & \textBF{0.0051} & 0.0185 & 0.0054 & 0.0477 \\ 
	    & & 14 & 0.0067 & \textBF{0.0131} & 0.0064 & 0.0440 & \textBF{0.0054} & 0.0200 & 0.0054 & 0.0440 \\ 
	    & & 15 & 0.0070 & \textBF{0.0041} & 0.0070 & 0.0275 & \textBF{0.0056} & 0.0365 & 0.0061 & 0.0365 \\ 
	    & & 16 & 0.0078 & \textBF{0.0050} & 0.0073 & 0.0230 & \textBF{0.0063} & 0.0230 & 0.0072 & 0.0230 \\ 
\cmidrule{2-11}
	    & & Mean & 0.0049 & 0.0205 & 0.0047 & 0.0407 & \textBF{0.0039} & \textBF{0.0159} & 0.0043 & 0.0201 \\ 
\bottomrule
\end{longtable}
\end{center}   

\vspace{-.6in}

\section{Annuity price calculation under an interest rate $\eta=3\%$}\label{sec:appendix_D}

In Table~\ref{tab:Appendix_4}, we display the annuity price calculation for female and male populations residing in Japan under a constant interest rate of $\eta=3\%$ in line with the US, UK and Australia. In contrast to $\eta=0.25\%$, the following annuity premiums are lower for each age and maturity period. 
\begin{table}[!htb]
\centering
\tabcolsep 0.07in
\caption{\small{Estimates of annuity prices with different ages and maturities ($T$) for female and male policyholders residing in Japan. These estimates are based on forecast mortality from 2023 to 2072 and a constant interest rate $\eta=3\%$. We consider only contracts with maturity so that age + maturity $\leq 110$.}\label{tab:Appendix_4}}
\begin{tabular}{@{}rrrrrrrrrrrrr@{}}
  \toprule
	& \multicolumn{6}{c}{Female} 	& \multicolumn{6}{c}{Male} \\	
Age & $T=5$ & 10 & 15 & 20 & 25 & 30 & 5 & 10 & 15 & 20 & 25 & 30 \\ 
  \midrule
  60 & 4.532 & 8.352 & 11.535 & 14.131 & 16.158 & 17.594 & 4.481 & 8.154 & 11.081 & 13.313 & 14.894 & 15.854 \\ 
  65 & 4.513 & 8.273 & 11.340 & 13.734 & 15.431 & 16.402 & 4.431 & 7.960 & 10.652 & 12.559 & 13.717 & 14.233 \\ 
  70 & 4.477 & 8.128 & 10.979 & 12.999 & 14.155 & 14.590 & 4.347 & 7.662 & 10.011 & 11.438 & 12.073 & 12.244 \\ 
  75 & 4.412 & 7.858 & 10.300 & 11.696 & 12.221 & 12.317 & 4.226 & 7.220 & 9.038 & 9.848 & 10.066 & 10.092 \\ 
  80 & 4.278 & 7.310 & 9.043 & 9.696 & 9.815 & 9.823 & 4.022 & 6.464 & 7.551 & 7.844 & 7.880 & 7.882 \\ 
  85 & 3.998 & 6.284 & 7.144 & 7.301 & 7.311 &  & 3.614 & 5.223 & 5.657 & 5.711 & 5.713 &  \\ 
  90 & 3.450 & 4.749 & 4.985 & 5.002 &  &  & 2.965 & 3.764 & 3.863 & 3.868 &  &  \\ 
  95 & 2.641 & 3.122 & 3.154 &  &  &  & 2.188 & 2.459 & 2.472 &  &  &  \\ 
  100 & 1.743 & 1.861 &  &  &  &  & 1.465 & 1.536 &  &  &  &  \\ 
  105 & 1.117 &  &  &  &  &  & 0.993 &  &  &  &  &  \\ 
   \bottomrule
\end{tabular}
\end{table}

\newpage
\bibliographystyle{agsm}
\bibliography{model_cdf.bib}

\end{document}